\documentclass[preprint2]{aastex701}

\begin{document}

\title{Pre-Supernova Eruptions Triggered by Sudden Energy Deposition in Low-Mass Core-Collapse Supernova Progenitors}

\author[orcid=0000-0001-6773-7830]{Shuai Zha}
\email[show]{zhashuai@ynao.ac.cn}  
\affiliation{International Centre of Supernovae (ICESUN), Yunnan Key Laboratory of Supernova Research, Yunnan Observatories, Chinese Academy of Sciences (CAS), Kunming 650216, People's Republic of China}
\author{Han Lin}
\email{1437047176@qq.com}  
\affiliation{International Centre of Supernovae (ICESUN), Yunnan Key Laboratory of Supernova Research, Yunnan Observatories, Chinese Academy of Sciences (CAS), Kunming 650216, People's Republic of China}
\author[orcid=0000-0001-5284-8001]{Xuefei Chen}
\email[show]{cxf@ynao.ac.cn}  
\affiliation{International Centre of Supernovae (ICESUN), Yunnan Key Laboratory of Supernova Research, Yunnan Observatories, Chinese Academy of Sciences (CAS), Kunming 650216, People's Republic of China}
\author[orcid=0000-0001-9204-7778]{Zhanwen Han}
\email[show]{zhanwenhan@ynao.ac.cn}  
\affiliation{International Centre of Supernovae (ICESUN), Yunnan Key Laboratory of Supernova Research, Yunnan Observatories, Chinese Academy of Sciences (CAS), Kunming 650216, People's Republic of China}

\begin{abstract}
In low-mass core-collapse supernova (CCSN) progenitors, nuclear burning beyond oxygen can become explosive under degenerate conditions, triggering eruptive mass loss before the final explosion. We investigate such pre-SN eruptions using \texttt{SNEC} hydrodynamic simulations and realistic stellar models, parameterizing the nuclear energy deposition as a fraction of the binding energy of the combined He layer and H-rich envelope. 
For the lowest-mass model (9\,$M_\sun$), the ejecta mass ($M_{\rm ej}$) scales with the energy gained by the H-rich envelope via a power law (index$\sim$3.5). Across 9-10\,$M_\sun$, this relation shows limited scatter within a factor of $\sim$2.6, enabling an estimation of the gained energy from $M_{\rm ej}$. The shock passage also flattens the bound envelope, which can affect the SN light curve morphology and provide another diagnostic for the eruption. 
Then, we compute the associated precursor light curves for the 9\,$M_\sun$ model with the multi-group radiative-transfer code \texttt{STELLA}.  These signals are typically faint, with bolometric luminosities of $\sim10^{39}$\,erg s$^{-1}$ lasting hundreds of days. Their cool black-body spectra make them brighter in the infrared, yet several magnitudes fainter than observed pre-SN precursors at the threshold for full envelope ejection. To aid future studies, we make our post-eruption stellar profiles and precursor light curves publicly available.
\end{abstract}

\keywords{\uat{Core-collapse supernovae}{304} --- \uat{Late stellar evolution}{911} --- \uat{Circumstellar matter}{241} --- \uat{Transient sources}{1851} }

\section{Introduction}
Core-collapse supernovae (CCSNe), the explosive deaths of massive stars with a zero-age main-sequence mass ($M_{\rm ZAMS}$) greater than $\sim$8-10\,$M_\sun$ (e.g. \citealt{2009ARA&A..47...63S,2013ApJ...765L..43I}), constitute a major class of research objects for wide-field time-domain astronomical surveys~(see e.g., \citealt{ZTF,LSST,ASASSN,ATLAS,PANSTARS,WFST,MEPHISTO}). Thanks to the high cadence of modern surveys and their rapid follow-up observations, a growing sample of hydrogen-rich CCSNe, i.e. Type II supernovae (SNe II), observed at early phases exhibits the features of interactions with dense circumstellar material (CSM) surrounding their progenitor stars (\citealt{2000ApJ...536..239L,2014Natur.509..471G,2015MNRAS.449.1876S,2016ApJ...818....3K,yaron2017,2018NatAs...2..808F,2021ApJ...912...46B,2023ApJ...952..119B}). This CSM directly manifests the mass-loss activities during the final evolutionary stages (the last months to years) of CCSN progenitors \citep{2017hsn..book..403S}, thereby serving as a powerful diagnostic for probing the final phases of massive stars immediately preceding their explosion.

Stars undergo mass loss throughout their lifetimes, and for single stars, stellar evolutionary models typically incorporate empirical prescriptions for steady-state, radiation- or dust-driven winds \citep{kippenhahn2012,langer2012}. However, \cite{2021ApJ...912...46B,2023ApJ...952..119B} inferred that $>$30\% of SN II progenitors—likely red supergiants (RSGs)—experienced enhanced mass loss immediately before explosion to form dense and confined CSM, with rates exceeding standard prescriptions. Several models have been proposed to resolve the tension without invoking significantly elevated mass loss rates, such as accelerating winds \citep{2010ASPC..425..181B,2017MNRAS.469L.108M}, a pre-explosion effervescent zone \citep{2021ApJ...906....1S}, or a dense chromosphere \citep{2017A&A...605A..83D,2024OJAp....7E..47F}. 
Another intriguing way to form the observed dense CSM is a brief episode of eruptive mass loss \citep{2010MNRAS.405.2113D,2012MNRAS.423L..92Q,2014ApJ...785...82S,2020ApJ...891L..32M,2020A&A...635A.127K,2020ApJ...900...99L,2021ApJ...908...23M,2022ApJ...936..114M,2023ApJ...945..104T,2024ApJ...974..270C}. The eruptive scenario finds observational support in the appearance of nearby SN IIP progenitors \citep{2020MNRAS.493..468D,2022MNRAS.517.1483D}. More direct evidence for such eruptions comes from outbursts preceding interacting SNe \citep{2009CBET.1928....1M,2011ApJ...732...32F,2013MNRAS.430.1801M,2014ApJ...789..104O,2015MNRAS.450..246B,2024A&A...684L..18B,2025A&A...701A..32P}, with 18 cases observed prior to Type IIn SNe by the Zwicky Transient Facility (ZTF) in 2018-2020 \citep{2021ApJ...907...99S}. However, with the notable exception of SN 2020tlf—the first normal Type IIP/L supernova with confirmed precursor emission \citep{2022ApJ...924...15J}—such activity appears to be typically too faint to observe in other normal SNe II.

The physical mechanism(s) responsible for such late-stage, pre-SN eruptions remain elusive, with candidates including internal gravity waves \citep{2012MNRAS.423L..92Q,2017MNRAS.470.1642F}, instabilities of advanced nuclear burning \citep{2011ApJ...733...78A,2014ApJ...785...82S}, opacity-driven super-Eddington luminosities \citep{2018Natur.561..498J,2024ApJ...974..270C}, electron-positron pair instabilities \citep{1967PhRvL..18..379B,2007Natur.450..390W}, and binary interactions \citep{2012ApJ...752L...2C,2019MNRAS.482.2277D}. Several numerical surveys have studied the eruptive scenario for CCSN progenitors with initial masses above 11\,$M_\sun$, exploring both sudden and continuous energy deposition at various locations within the star (see e.g., \citealt{2010MNRAS.405.2113D,2020ApJ...891L..32M,2023ApJ...945..104T}). \cite{2019MNRAS.485..988O} has explored such eruptions in very massive stars (100\,$M_\sun$) in the context of Luminous Blue Variable stars. To complement these efforts, we conduct a systematic study of low-mass CCSN progenitors with $M_{\rm ZAMS}$ between 9 and 10\,$M_\sun$. Despite the narrow mass range, such stars are relatively abundant, constituting $\sim$22\,\% of all SN II progenitors under the assumption of a bottom-heavy, power-law initial mass function  (slope $\alpha=-2.35$, \citealt{1955ApJ...121..161S}) and an upper mass limit of 18\,$M_{\sun}$ \citep{2015PASA...32...16S}. They are the leading progenitor channel for low-luminosity SNe II \citep{2005MNRAS.364L..33M,2012AJ....143...19V,2025PASP..137d4203D,2025arXiv250620068D,2025MNRAS.540.2591L}.

Theoretical stellar evolution models are scarcer for low-mass CCSN progenitors than for their more massive counterparts, largely due to the numerical challenges of simulating off-center nuclear flashes and flames in partially degenerate cores during advanced burning stages (beyond carbon burning; \citealt{1980tsup.work...96W,1986PrPNP..17..267N,2013ApJ...772..150J,2014ApJ...797...83J,2015ApJ...810...34W,limongi2024}). In particular, explosive off-center nuclear flashes in advanced burning shells (e.g. silicon) provide a unique physical mechanism that releases energy in a short period and triggers pre-SN eruptions of progenitor stars with $M_{\rm ZAMS}\sim$9-10\,$M_\sun$, potentially giving rise to faint IIP-like precursor transients \citep{1979BAAS...11Q.724W,2015ApJ...810...34W}. The timing, location, and intensity of this violent energy release—and its subsequent deposition in the outer mantle—are highly uncertain, as they depend on the detailed ignition conditions of the nuclear flashes. Here, we conduct a parameter study of this eruptive phenomenon to quantify the relationship between the amount of energy deposition and the resulting mass ejection and precursor signals. 

Our paper is organized as follows. Section~\ref{sec:erupt} details our investigation of the eruptive mass loss through non-radiative hydrodynamics simulations with parameterized energy deposition. We detail our simulation methodology and progenitor models (\S~\ref{ssec:method}), present a detailed analysis of the $9.0$\,$M_\sun$ case (\S~\ref{ssec:9.0}), and explore the progenitor dependence (\S~\ref{ssec:progs}). Section~\ref{sec:lcs} presents the associated precursor signals generated with the multi-group radiative-hydrodynamics code \texttt{STELLA}. We discuss the caveats of our study in Section~\ref{sec:discuss} and conclude our findings in Section~\ref{sec:conclu}.

\section{Eruption simulation \label{sec:erupt}}

\subsection{Simulation method and progenitors \label{ssec:method}}

In our pre-SN eruption scenario, the sudden energy deposition in deep layers due to explosive nuclear flashes launches a shock wave that propagates outward, unbinding part of the stellar envelope. We model the shock propagation and the subsequent mass ejection in spherical symmetry using the Lagrangian hydrodynamics code \texttt{SNEC} \citep{2015ApJ...814...63M}. To efficiently track the mass outflow over a sufficiently long timescale (up to 10 years), we operate \texttt{SNEC} in its non-radiative mode. The code works with the Paczynski equation of state (EOS; \citealt{1983ApJ...267..315P}) for a mixture of ions, electrons, and photons, which is further complemented with the partial ionization of H and He through a Saha solver \citep{2000JPhD...33..977Z,timmes1999,timmes2000}. 

In most of our simulations, we start by removing the progenitor's C-O core and depositing an energy $E_{\rm dep}$ at the base of the He layer. In the scenario of advanced nuclear flashes, energy deposition is expected to originate from a deeper region within the core. However, due to significant uncertainty in its exact location, we adopt the base of the He layer as a uniform treatment across our models. This energy is spread over 0.1\,$M_\sun$ within $10^3$\,s, which is much shorter than the dynamical timescale of the envelope ($\mathcal{O}(10^6)$\,s). The computational domain is discretized using a non-uniform grid consisting of 2000 zones across both the He layer and H-rich envelope. The mass resolution, defined as $\mathrm{d}q=\mathrm{d}M/M_{\rm tot}$, is $2\times10^{-3}$ at the midpoint (zone 1000), and it decreases proportionally to $2\times10^{-4}$ at the innermost zone and $2\times10^{-6}$ at the outermost zone. We have confirmed that using a grid with 2.5$\times$ finer resolution results in nearly identical ejected masses across the explored range of $E_{\rm dep}$, compared to simulations conducted with our standard resolution.

\begin{deluxetable*}{rrrrrrr}[ht!]
%\digitalasset
\tablewidth{0pt}
\tablecaption{Properties of the stellar models \label{tab:prog}}
\tablehead{
\colhead{$M_{\rm ZAMS}$} & \colhead{$M_{\rm prog}$} & \colhead{$R_{\rm prog}$} & \colhead{$M_{\rm He}$}  & \colhead{$M_{\rm Henv}$} & \colhead{$E_{\rm bind,tot}$} & \colhead{$E_{\rm bind,H}$} \\
\colhead{($M_\odot$)} & \colhead{($M_\odot$)} & \colhead{($R_\sun$)} & \colhead{($M_\odot$)}  & \colhead{($M_\odot$)} & \colhead{(10$^{48}$\,erg)} & \colhead{(10$^{47}$\,erg)}
}
\startdata
9.0  &  8.75  & 412  & 0.17 & 7.18 & 2.17  & 0.98 \\
9.25 &  8.98  & 404  & 0.33 & 7.20 & 4.92  & 1.31 \\
9.5  &  9.21  & 412  & 0.54 & 7.17 & 7.43  & 1.52 \\
9.75 &  9.45  & 445  & 0.75 & 7.14 & 9.37  & 1.46 \\
10.0 &  9.68  & 513  & 0.87 & 7.20 & 9.37  & 0.77 \\
10.5 &  10.20 & 542  & 0.98 & 7.52 & 12.44 & 1.12 \\
11.0 &  10.70 & 569  & 1.03 & 7.87 & 16.74 & 1.12 \\
12.0 &  10.91 & 635  & 1.04 & 7.78 & 27.21 & 0.88 \\
\enddata
\tablecomments{$M_{\rm ZAMS}$ is the zero-age main-sequence mass. Other quantities are at the terminal of their standard evolution: $M_{\rm prog}$ and $R_{\rm prog}$ are the stellar mass and radius; $M_{\rm He}$ and $M_{\rm Henv}$ are the masses of He layer and H-rich envelope; $E_{\rm bind,tot}$ and $E_{\rm bind,H}$ are the binding energy for the combined He layer and H-rich envelope, and for the H-rich envelope only, respectively. In this paper, we performed eruption simulations using 5 models with $M_{\rm ZAMS}=9$-10\,$M_\sun$.}
\end{deluxetable*}

\begin{figure*}[t!]
    \centering
    \includegraphics[width=0.97\textwidth]{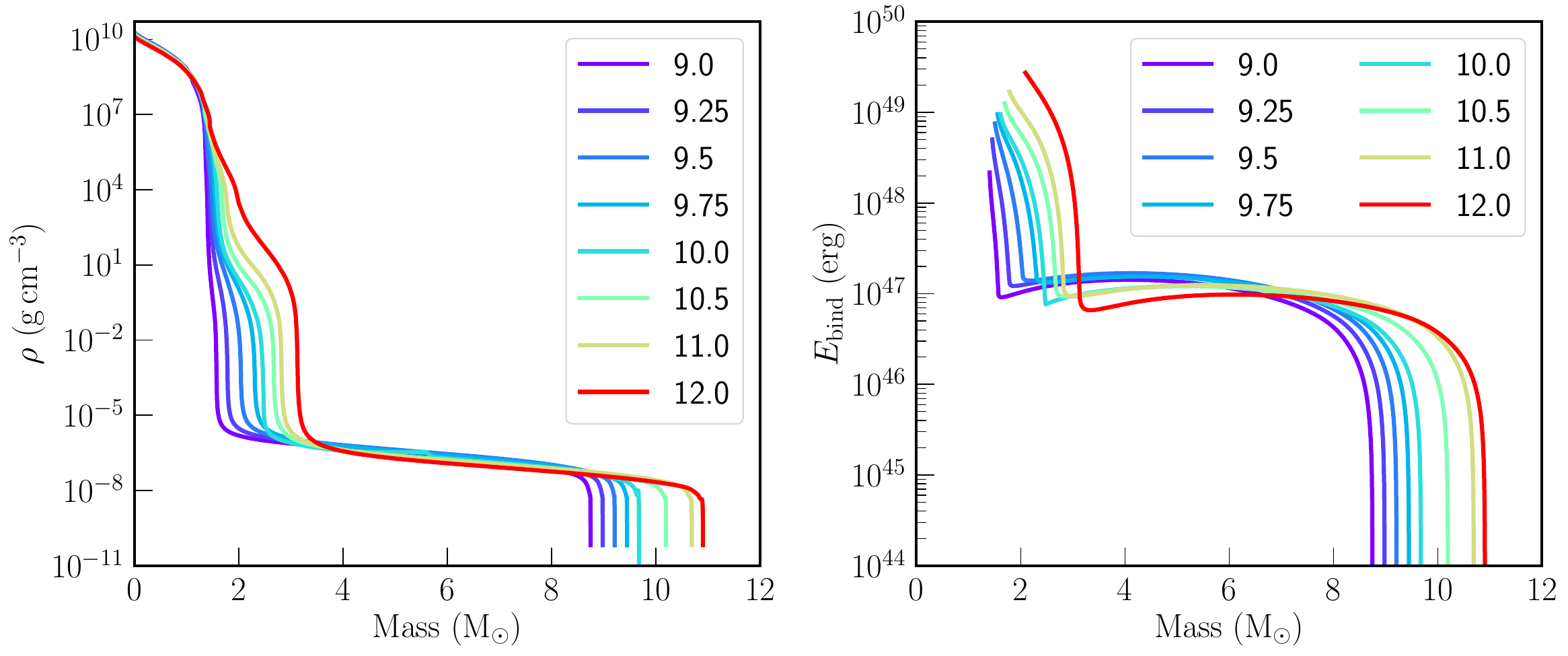}
    \caption{Profiles of density (left panel) and binding energy (right panel, computed as Eq.~\ref{eq:eb}) as a function of mass coordinate for our employed progenitor models ($M_{\rm ZAMS}=9$-10\,$M_\sun$) alongside three more massive ones up to 12\,$M_\sun$ for comparison. All models are from \cite{2016ApJ...821...38S}.}
    \label{fig:progs}
\end{figure*}

We use 5 solar-metallicity CCSN progenitor models with $M_{\rm ZAMS}=$9-10\,$M_\sun$ in increments of $0.25\,M_\sun$ evolved in the \texttt{Kepler} code (see Table~\ref{tab:prog}; \citealt{2016ApJ...821...38S})\footnote{This set of progenitor models was computed using erroneous pair neutrino loss rates. This mainly affects the core structure of models with $M_{\rm ZAMS}\ge20\,M_\sun$ and has negligible impact on the envelope structure for low-mass models \citep{2018ApJ...860...93S}.} as our initial conditions. These non-rotating and single-star models are widely employed in CCSN simulations (e.g., \citealt{2020ApJ...890..127C,2020MNRAS.496.2039S,2024ApJ...969...74W}) and light-curve modeling of SNe II \citep{2022ApJ...934...67B,2022MNRAS.514.4173K,2023ApJ...952..155Z,2023PASJ...75..634M,2024ApJ...970..163S}. The models were evolved until the immediate onset of core collapse, defined as the time when the maximum infall velocity exceeds 900\,km\,s$^{-1}$ in the iron core. Beyond core O burning, an RSG's envelope is effectively ``frozen" in the standard single-star evolution \citep{2002RvMP...74.1015W,langer2012}. Therefore, we adopt the pre-collapse envelope structure as a valid approximation for our purpose, following previous works \citep{2010MNRAS.405.2113D,2023ApJ...945..104T}. 

Table~\ref{tab:prog} summarizes the global properties of these stellar models, alongside 3 more massive cases up to $M_{\rm ZAMS}=12$\,$M_\sun$. The $M_{\rm ZAMS}=9$-10\,$M_\sun$ models lose 0.25-0.30\,$M_\sun$ in total through winds and retain H-rich envelopes of $\sim7.2\,M_\sun$. We list the binding energy, computed as 
\begin{equation} \label{eq:eb}
    E_{\rm bind}(M_r) = {\int_{M_r}^{M_\mathrm{max}}} {(GM'_r/r-\epsilon) \mathrm{d}M'_r},
\end{equation}
for the combined He layer and the H-rich envelope ($E_{\rm bind,tot}$), and for the H-rich envelope alone ($E_{\rm bind,H}$). Here, $\epsilon$ is the specific internal energy taken from \texttt{SNEC}. The CO/He and He/H interfaces are defined where the He and H fractions drop below 0.5, respectively. $E_{\rm bind,H}$ is of order $10^{47}$\,erg for all 9-12\,$M_\sun$ models, while $E_{\rm bind,tot}$ for the 9\,$M_\sun$ model is an order of magnitude smaller than that for the 12\,$M_\sun$ model and it differs by a factor of $\sim4$ within $M_{\rm ZAMS}=$9-10\,$M_\sun$. We note that there is a $\sim5$\% difference in the resulting $E_{\rm bind}$ of \texttt{SNEC} compared to original values of \texttt{Kepler}, due to the difference in their employed equation of state. We use the \texttt{SNEC} values as the reference energy to ensure consistency with the hydrodynamics simulations. In addition, the given progenitor models feature layer interfaces that span a mass range of a few $10^{-3}\,M_\sun$. This smearing introduces uncertainties of several percent in $E_{\rm bind,H}$ and $\sim$10\% in $E_{\rm bind,tot}$. Therefore, any correlation established using the absolute value of $E_{\rm dep}$ is sensitive to these model variations, while this dependence is greatly mitigated by using fractional energy deposition ($E_{\rm dep}/E_{\rm bind}$) as the control parameter.

Figure~\ref{fig:progs} further shows the density and binding energy profiles as a function of the mass coordinate for these models. The \texttt{Kepler} progenitor models exhibit negative specific binding energy at the bottom of the H-rich envelope, indicating that this region is more easily ejected. This leads to a distinctive mass ejection behavior for energy deposition near the threshold of full envelope ejection. We have checked that \texttt{MESA} CCSN progenitor models (e.g., evolved with example\_ccsn\_IIp in the test suite of \texttt{MESA} revision 12778, \citealt{MESA18,2023ApJ...945..104T}), do not display this feature.  This difference between stellar evolution codes merits further investigation. Finally, uncertainties in the treatment of stellar outer boundary condition \citep{2018ApJ...860...93S} and envelope convection (e.g., the mixing-length parameter and convective overshooting, \citealt{1987A&A...182..243M,2013MNRAS.433.1745D}) can substantially alter the envelope structure. A comprehensive assessment of these effects is beyond the scope of this work. 

In this work, we mainly discuss results with $E_{\rm dep}$ in the range of 0.5-1.0$\times E_{\rm bind,tot}$. As we will demonstrate, the lower limit of 0.5 corresponds to a non-negligible fraction of mass ejection ($\sim10^{-4}$), whereas the upper limit corresponds to the full envelope ejection. This energy deposition can be related to an explosive nuclear flash of O or Si burning, which releases energy in a dynamical timescale during the late-stage evolution of low-mass CCSN progenitors \citep{2015ApJ...810...34W}. For instance, converting 0.01\,$M_{\sun}$ of O to Si releases $9.06\times10^{48}$\,erg, and converting 0.01\,$M_{\sun}$ of Si to Ni releases $3.74\times10^{48}$\,erg. These energy yields are on the same order as $E_{\rm bind,tot}$, making our chosen $E_{\rm dep}$ physically motivated. 

\subsection{The 9.0\,$M_\sun$ model \label{ssec:9.0}}

\subsubsection{Structural response}

\begin{figure*}[t]
    \centering
    \includegraphics[width=0.97\textwidth]{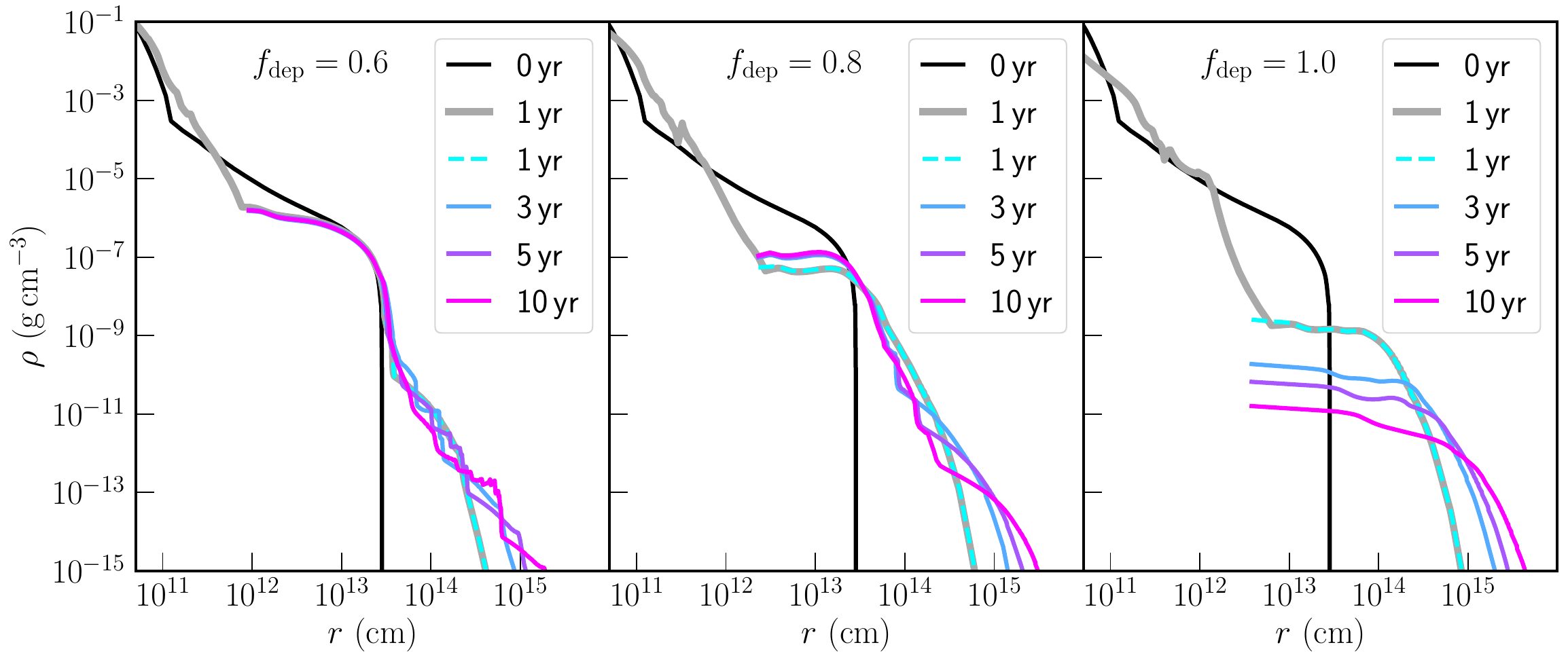}
    \caption{Density profiles as a function of radius at selected times for 3 eruption simulations using the 9\,$M_\sun$ model. Fractional energy deposition ($f_{\rm dep}\equiv E_{\rm dep}/E_{\rm bind,tot}$) is 0.6, 0.8, and 1.0 from left to right.  The solid gray and dashed cyan lines represent the results at 1 year for the full and H-only simulations, respectively. After 1 year, we only plot the results for the long-term H-only simulations. }
    \label{fig:rho}
\end{figure*}

We first present a detailed analysis of the eruption simulations for the lowest-mass model in our set, i.e. 9.0\,$M_\sun$. This model has the smallest values for the He-layer mass and $E_{\rm bind,tot}$. Figure~\ref{fig:rho} shows density profiles at selected times for 3 different fractional energy depositions, $f_{\rm dep}\equiv E_{\rm dep}/E_{\rm bind, tot}=0.6,~0.8$ and 1.0. For these cases, the shock generated by the sudden energy deposition breaks out of the stellar surface at $\sim$43, 33 and 25 days, respectively, with a shorter breakout time as $f_{\rm dep}$ increases. For sub-$E_{\rm bind,tot}$ energy deposition ($f_{\rm dep}<1$), the He layer remains nearly static after the shock passes through it. This leads to prohibitively small timesteps ($\mathcal{O}$(s)) in the simulation due to the high sound speed in the He layer. To mitigate this, we further remove the He layer at 80 days post deposition, after which the shock has broken out of the stellar surface for all simulated cases with $f_{\rm dep}\ge0.5$, and continue the simulation with only the H-rich envelope. We refer to these as ``H-only" simulations in contrast to the ``full" simulations that include both the He layer and the H-rich envelope, which were run for at least 1 year. As shown in each panel of Figure~\ref{fig:rho}, the solid gray and dashed cyan curves represent the density profiles of the full and H-only simulations 1 year after energy deposition, which are nearly identical for the overlapping regions.

Compared to the initial profile, the inner region of the evolved envelope (approximately within $R_{\rm prog}$) becomes progressively thinner and flatter as $f_{\rm dep}$ increases. This structural change has been suggested by \cite{2019ApJ...877...92O} and it will have a direct impact on the light curve of SNe II, such as making it more luminous and bluer \citep{2013MNRAS.433.1745D,2023ApJ...952..155Z}. \cite{2021MNRAS.500.1889O} proposed that this expansion of the H-rich envelope can explain the enhanced luminosity of some bright SNe IIP, e.g. SN 2009 kf. In the $f_{\rm dep}=$0.6 case, the envelope separates into two distinct parts slightly above the original stellar radius $R_{\rm prog}$ (the sharp density drop in the black curve). The inner region remains almost static after 1 year, while the outer region up to $\sim10^{15}$\,cm develops discontinuous features that likely lead to hydrodynamic instabilities and inhomogeneities in the CSM \citep{2022ApJ...936...28T}. For $f_{\rm dep}=$0.8, the inner region contracts slightly before stabilizing after 3 years. This static part extends to $\sim10^{14}$\,cm (or $\sim3.5\times R_{\rm prog}$), with a density slope of $\sim4.9$ outside $\sim R_{\rm prog}$, while the outer part expands homologously ($v_r \propto r$). In stark contrast, the $f_{\rm dep}=1.0$ case results in the complete, homologous ejection of the entire H-rich envelope. If the final CCSN explosion occurs after a significant delay, it may not display the usual H recombination signature of SNe II.

\begin{figure*}[t]
    \centering
    \includegraphics[width=0.97\textwidth]{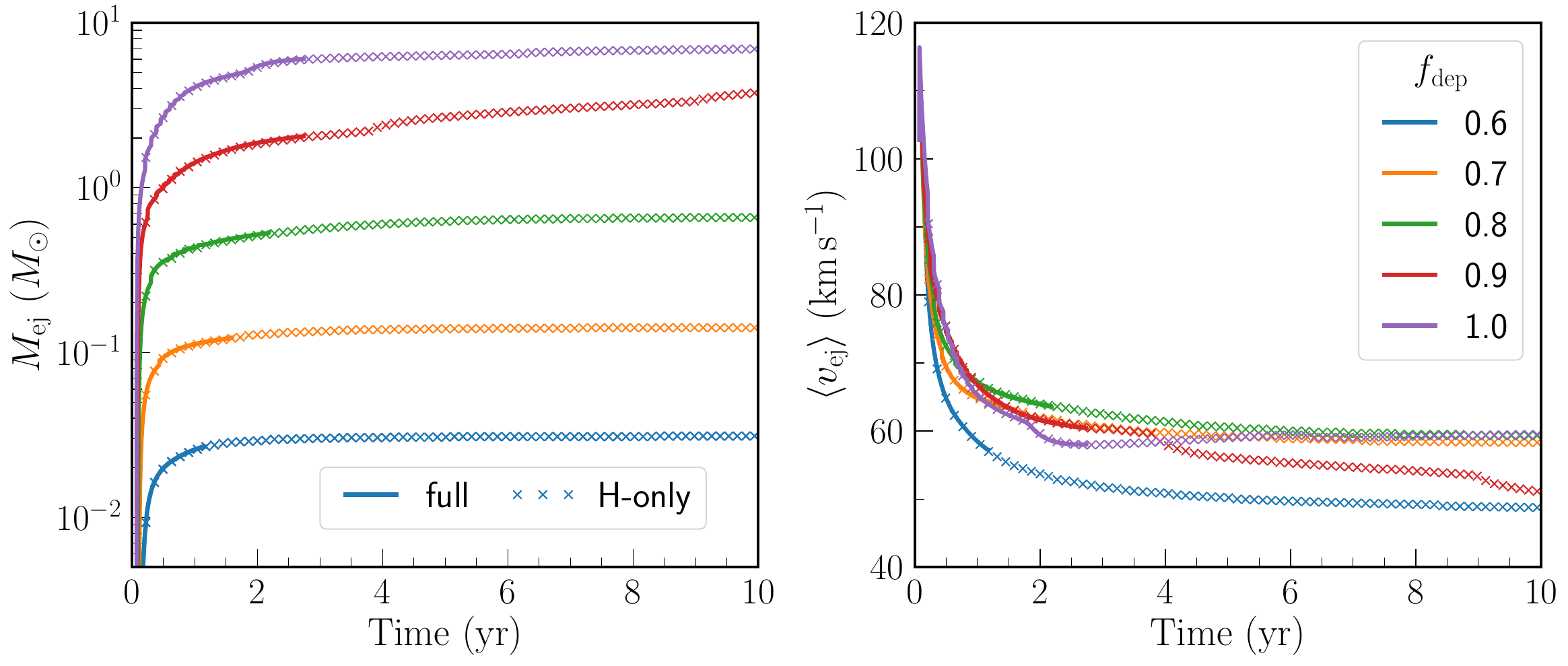}
    \caption{Time evolution of the ejecta mass ($M_{\rm ej}$, left panel) and mean velocity ($\langle v_{\rm ej} \rangle$, right panel) for selected fractional energy deposition ($f_{\rm dep}=E_{\rm dep}/E_{\rm bind,tot}$). Solid curves represent the full simulations including the He layer and H-rich envelope, while ``$\times$" symbols show simulations where the He layer was removed at 80\,days.}
    \label{fig:ej_vs_t}
\end{figure*}

\subsubsection{Mass ejection process}

A key question concerns how the total mass ejected and the kinematics of the ejection process depend on $f_{\rm dep}$. To investigate this, we plot the ejecta mass ($M_{\rm ej}$) and its mean velocity ($\langle v_{\rm ej} \rangle$) as a function of time for $0.6\le f_{\rm dep}\le1.0$ in Figure~\ref{fig:ej_vs_t}. Here, we adopt a conservative criterion for the ejecta that requires the outflow velocity to exceed the local escape velocity, $v_{\rm esc}=\sqrt{2GM_r/r}$. A comparison of the full (solid curves) and H-only (``x" symbols) simulations shows close agreement for the overlapping periods in both panels, further demonstrating that removing the quasi-static He layer is a valid approach. The results reveal clear trends in mass ejection kinematics. $M_{\rm ej}$ reaches its saturation after $\sim 3$ years for all values of $f_{\rm dep}$ except 0.9. The saturated $M_{\rm ej}$ increases with $f_{\rm dep}$ as expected. Concurrently, $\langle v_{\rm ej} \rangle$ decreases from an initial maximum of $\sim120$\,km\,s$^{-1}$, $\sim$30\% above the original RSG's surface escape velocity ($\sim90$\,km\,s$^{-1}$), to a final static value of $\sim50$\,km\,s$^{-1}$ for $f_{\rm dep}=0.6$ and $\sim60$\,km\,s$^{-1}$ for all other $f_{\rm dep}$ values except 0.9. 

\begin{figure}
    \centering
    \includegraphics[width=0.97\linewidth]{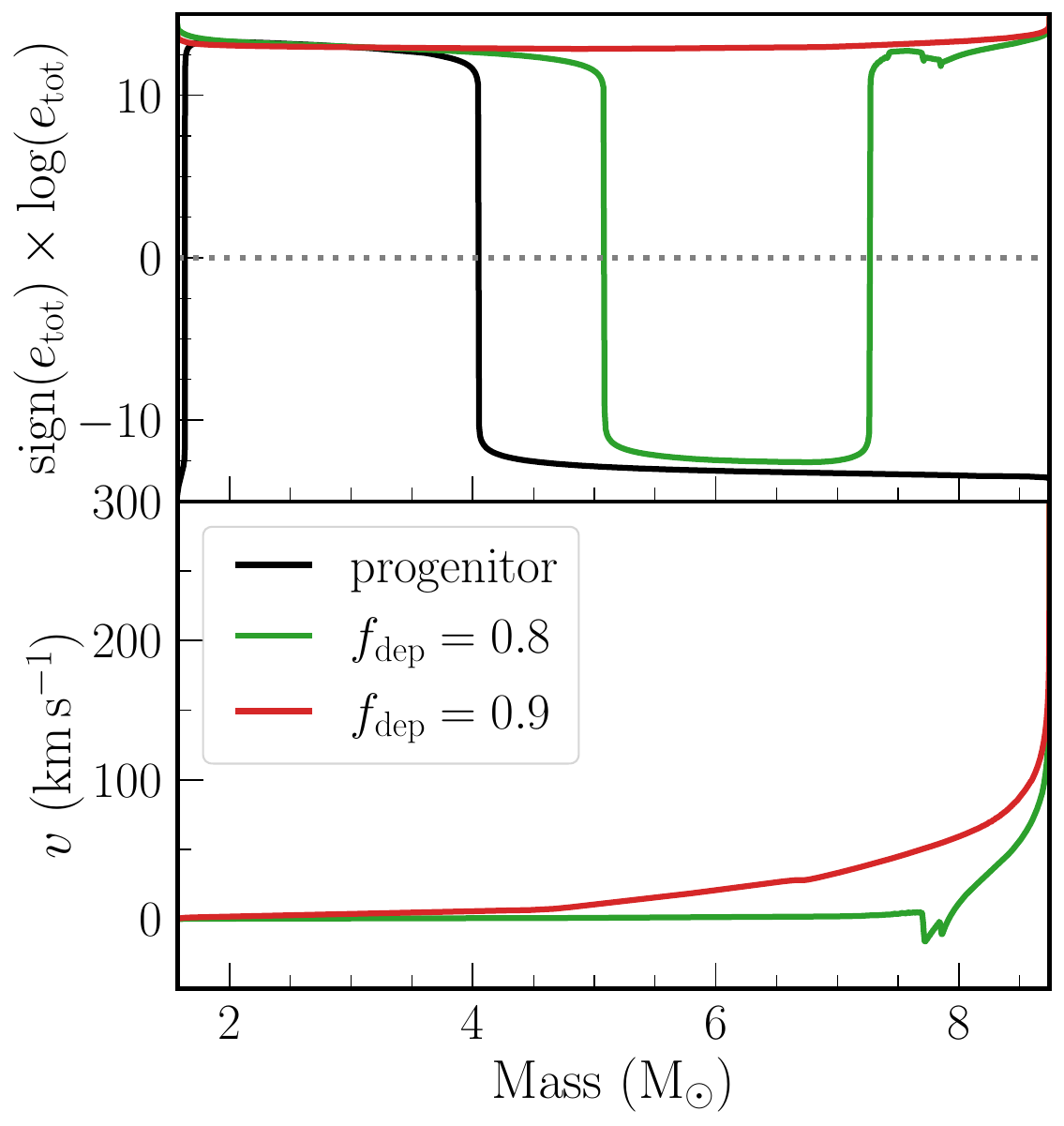}
    \caption{Profiles of total specific energy ($e_{\rm tot}$; top) and velocity (bottom) as a function of mass coordinate for the progenitor model (black) and for $f_{\rm dep} = 0.8$ (green) and 0.9 (red) at 5 years post energy deposition. A signed logarithmic scale is used for $e_{\rm tot}$ to enhance clarity.}
    \label{fig:0.8vs0.9}
\end{figure}

For $f_{\rm dep}=0.9$, both $M_{\rm ej}$ and $\langle v_{\rm ej} \rangle$ exhibit correlated long-term trends, with the mass increasing and the mean velocity decreasing steadily. Superimposed on this trend are concurrent features: $M_{\rm ej}$ shows small jumps at $\sim$4 and 9 years, while $\langle v_{\rm ej} \rangle$ exhibits clear decreases at the same times. To diagnose this distinct behavior, we compare the total specific energy ($e_{\rm tot}$, sum of gravitational, internal, and kinetic) and velocity profiles for models with $f_{\rm dep}$=0.8 and 0.9 at 5 years post energy deposition in Figure~\ref{fig:0.8vs0.9}. For clarity, we plot the signed logarithm of $e_{\rm tot}$. As already noted in \S~\ref{ssec:method}, the progenitor model has a region of positive $e_{\rm tot}$ at the base of the H-rich envelope (the black curve in Figure~\ref{fig:0.8vs0.9}). For the $f_{\rm dep}$=0.8 case, a zone of negative $e_{\rm tot}$ and therefore a strictly bound region persists between the ejecta and the inner part of the envelope with positive $e_{\rm tot}$. In contrast, for $f_{\rm dep}$=0.9, the entire envelope gains positive $e_{\rm tot}$, facilitating more mass ejection but with the inner ejecta expanding at lower velocities.  For \texttt{MESA}-like progenitors where $e_{\rm tot}$ is negative throughout the H-rich envelope (e.g., \citealt{2023ApJ...945..104T}), we anticipate more self-similar kinematics across the parameter space of $f_{\rm dep}$ with sub-$E_{\rm bind,tot}$ deposition. Nevertheless, such a distinct behavior associated with the $e_{\rm tot}$ profile warrants further investigation in the modeling of CCSN progenitors.

\begin{figure*}
    \centering
    \includegraphics[width=0.475\textwidth]{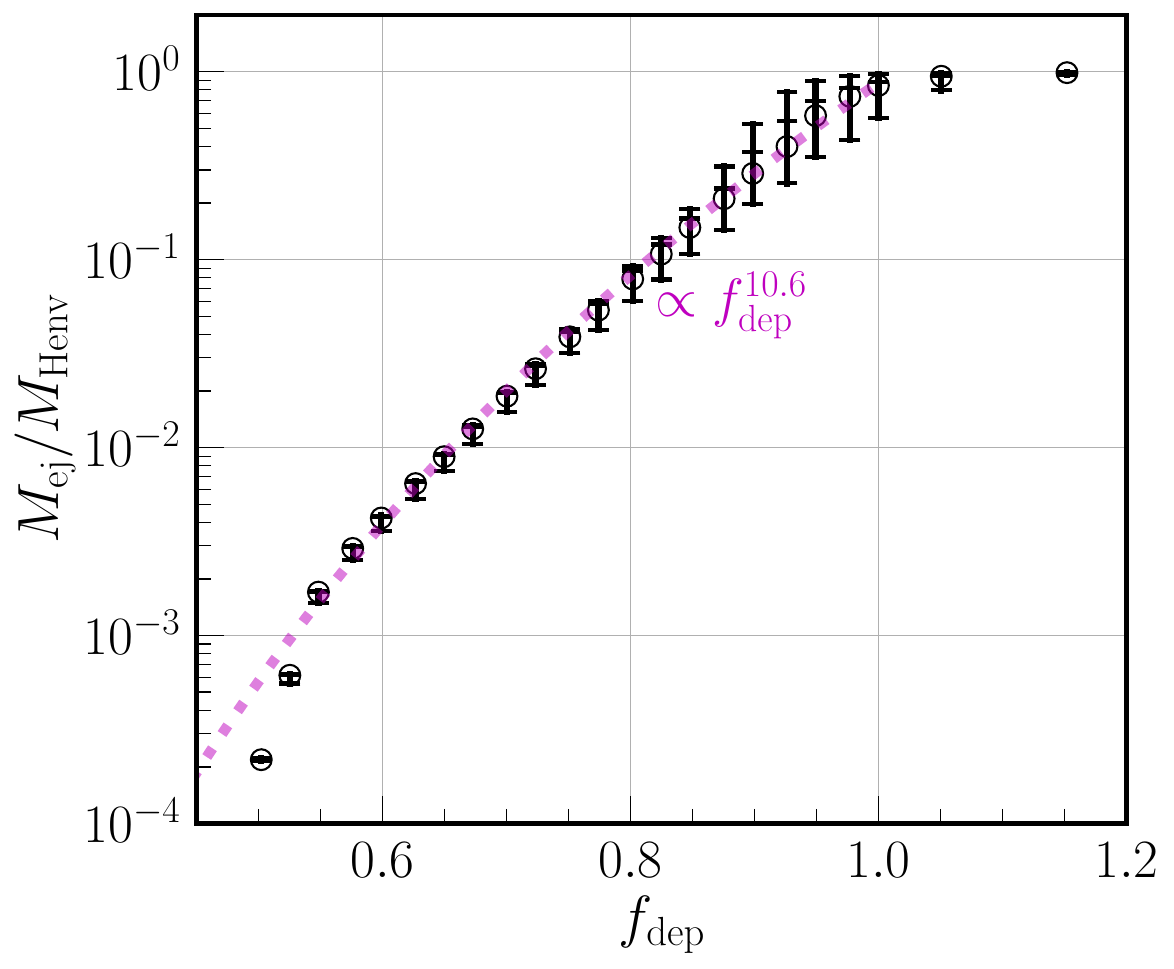}
    \includegraphics[width=0.475\textwidth]{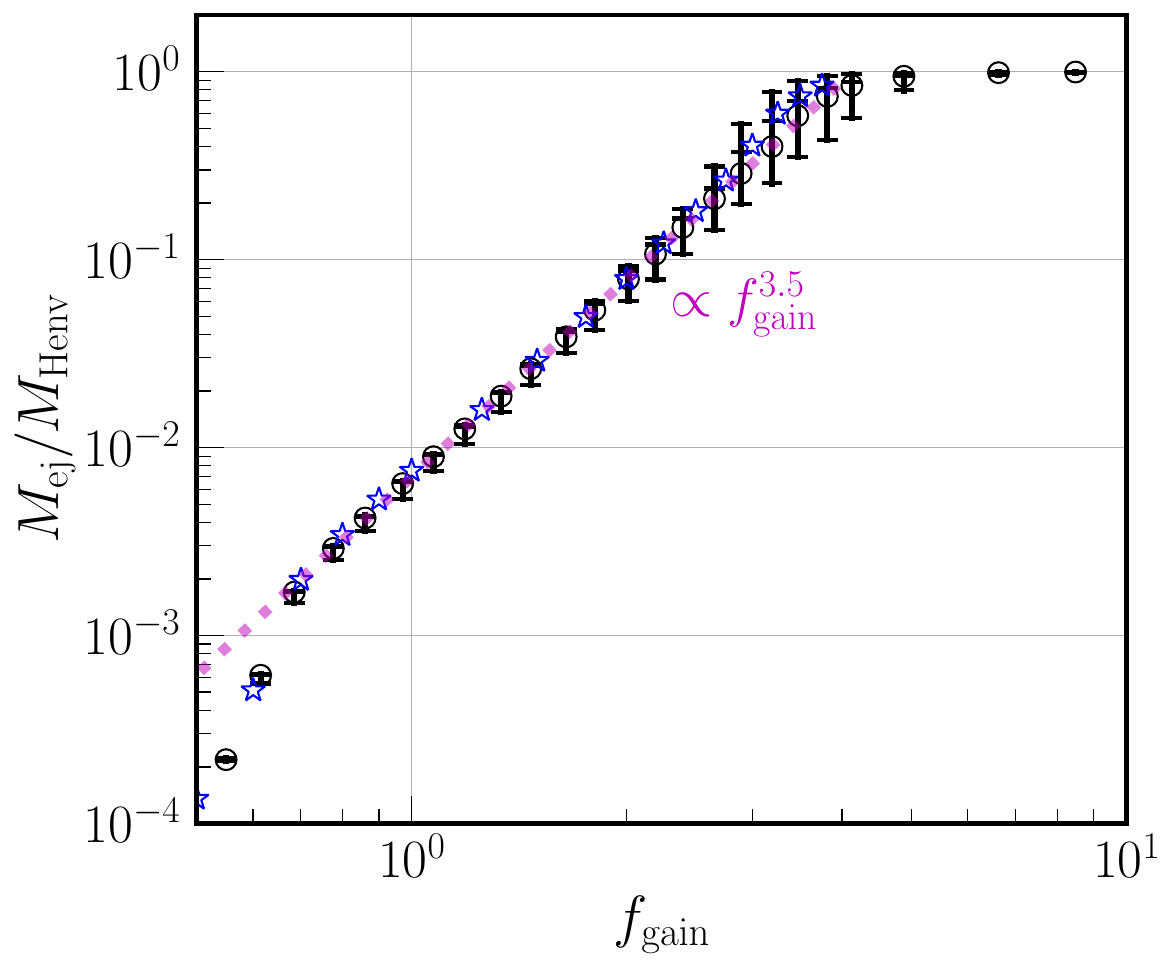}
    \caption{The fraction of ejected mass relative to the mass of the H-rich envelope ($M_{\rm ej}/M_{\rm Henv}$) as a function of fractional energy deposition ($f_{\rm dep}$, left panel) and fractional energy gained by the H-rich envelope ($f_{\rm gain}$, right panel). Black  bars show the results at 1, 5 and 10 years (from bottom to top) while the open circles mark the results at 3 years after energy deposition. In each panel, the magenta curves is a power-law fit to the data. In the right panel, the blue stars show the results with fractional energy deposited at the base of the H-rich envelope which show excellent agreement with the main results.}
    \label{fig:mej_vs_f}
\end{figure*}

\subsubsection{$M_{\rm ej}$-$f_{\rm dep}$ relation}

Lastly, we quantify the relationship between $M_{\rm ej}$ and $f_{\rm dep}$ for the $9.0\,M_\sun$ model using a suite of simulations with $f_{\rm dep}$ ranging from 0.5 and 1.0 in increments of 0.025. In the left panel of Figure~\ref{fig:mej_vs_f}, black bars show $M_{\rm ej}$ at 1, 5 and 10 years post energy deposition from bottom to top, and open circles mark the data at 3 years. We find three distinct regions for mass ejection: for $f_{\rm dep}\lesssim0.85$, $M_{\rm ej}$ saturates within 3 years; for $f_{\rm dep}\gtrsim1.0$, the entire H-rich envelope is ejected within 3 years. A portion of the He layer can also be stripped for $f_{\rm dep}\gtrsim1.0$. In the intermediate regime ($0.85\lesssim f_{\rm dep}\lesssim 1.0$), $M_{\rm ej}$ increases over the long term, with nearly the entire H-rich envelope ejected by 10 years for $f_{\rm dep}\gtrsim0.95$. A power-law fit to the 3-year data (open circles) yields $M_{\rm ej}\propto f_{\rm dep}^{10.6}$, indicating a very sensitive dependence. \cite{2020A&A...635A.127K} have reported a similarly quick increase in $M_{\rm ej}$ with $f_{\rm dep}$ for more massive RSGs and yellow supergiants. 

Recognizing that only the energy gained by the H-rich envelope ($E_{\rm gain}$) drives the final eruptive mass loss, we further examine the dependence of mass ejection on $E_{\rm gain}$. The right panel of Figure~\ref{fig:mej_vs_f} shows $M_{\rm ej}$ as a function of the fractional energy gain, $f_{\rm gain}=E_{\rm gain}/E_{\rm bind,H}$. The entire H-rich envelope is ejected for $f_{\rm gain}\gtrsim4$. The relation is well-described by a power law with an index of $\sim3.5$ (dashed magenta curve), which is consistent with values found in simulations with polytropic stars \citep{2021MNRAS.501.4266L,2024ApJ...967...33C}. Specifically, \cite{2021MNRAS.501.4266L} obtained indices of 1.79-4.17 for a polytropic index $n$ of 1.5, closely resembling our RSG envelope. We further validate this relation using a separate simulation set where energy is deposited directly at the base of the H-rich envelope (blue open stars). In this configuration, $f_{\rm gain}$ is equivalent to $f_{\rm dep}$. The excellent agreement confirms that the mass eruption is primarily controlled by $E_{\rm gain}$. Therefore, since a significant portion of the energy at deeper layers is consumed before reaching the envelope, the ejecta mass directly reflects $E_{\rm gain}$ rather than the total $E_{\rm dep}$. Inferring the true energy deposition and its underlying physical mechanism requires a comprehensive analysis that incorporates energy release into realistic stellar evolutionary models.

\subsection{Progenitor dependence \label{ssec:progs}}

\begin{figure*}
    \centering
    \includegraphics[width=0.475\linewidth]{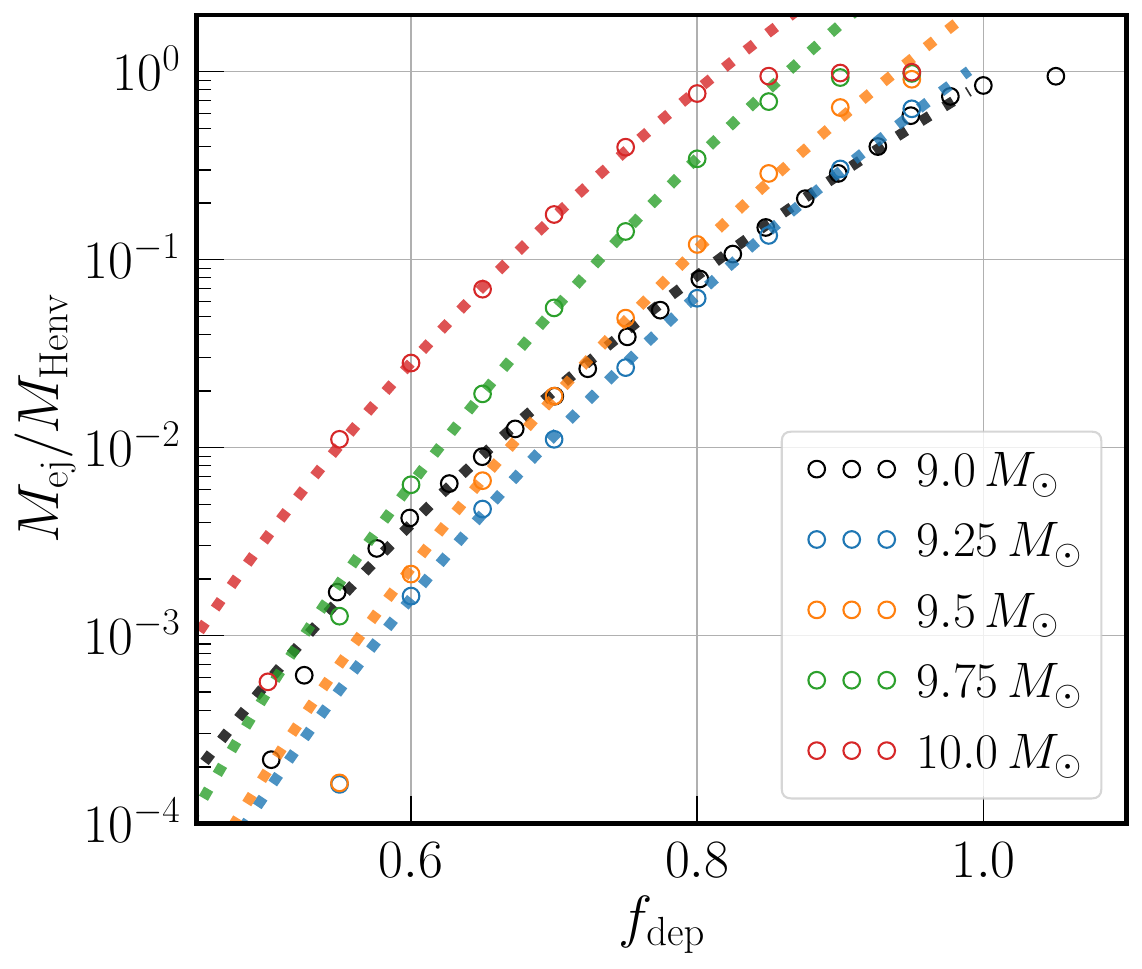}
    \includegraphics[width=0.475\linewidth]{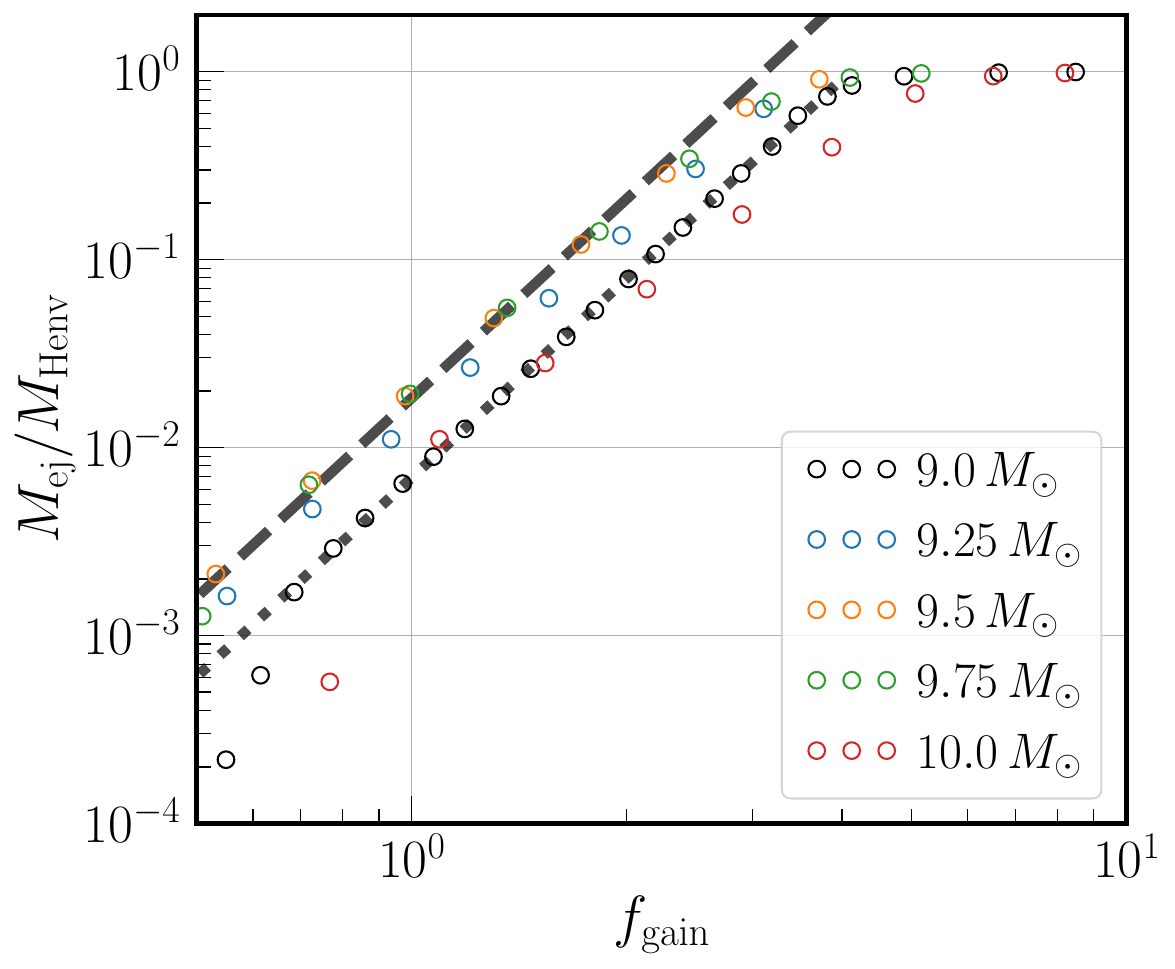}
    \caption{Same as Figure~\ref{fig:mej_vs_f}, but for the progenitor dependence with all data measured at 3 years post energy deposition. In the left panel, the dotted curves are power-law fits to the data with indices of $\sim$10.6-13.9. We fit the data over $f_{\rm dep}=0.55$-1.0 for models with $M_{\rm ZAMS}\le9.5\,M_\sun$, and over $f_{\rm dep}=0.55$-0.8 for the two more massive models. In the right panel, the black dotted curve shows the power-law fit for the 9.0\,$M_\sun$ model, and the dashed curve is the same fit scaled upward by a factor of 2.6.}
    \label{fig:mej_prog}
\end{figure*}

We explore the progenitor dependence within $M_{\rm ZAMS}=$9-10\,$M_\sun$, scanning $f_{\rm dep}$ from 0.5 to 1.0 in increments of 0.05. Figure~\ref{fig:mej_prog} shows the resulting ejected mass ($M_{\rm ej}$, evaluated at 3 years post energy deposition) as a function of $M_{\rm ZAMS}$ and energy deposition. The left panel shows that the $M_{\rm ej}$-$f_{\rm dep}$ relation is tightly clustered for $M_{\rm ZAMS}\le9.5$\,$M_\sun$ but exhibits significant deviation for more massive models. Even within this narrow mass range of 1\,$M_\sun$, the ejected fraction ($M_{\rm ej}/M_{\rm Henv}$) can vary by a factor of $\sim$10 at fixed $f_{\rm dep}$. The dotted curves show power-law fits to these relations for each stellar model with the same color. For the 2 more massive models (9.75 and 10.0\,$M_\sun$), we limit the fit to $f_{\rm dep}\le0.8$, as the entire H-rich envelope is ejected at a smaller value of $f_{\rm dep}$ ($\sim0.85$). The power-law index lies in the range of $\sim10.6$ to $13.9$ without a monotonic trend with $M_{\rm ZAMS}$. 

The right panel further illustrates the relation between $M_{\rm ej}$ and $f_{\rm gain}$. The dotted and dashed curves represent the fit for the 9\,$M_\sun$ model and the fit scaled upward by a factor of 2.6, respectively. The tight clustering in this plane demonstrates the potential to constrain the energy gained by the H-rich envelope by $M_{\rm ej}$ during such a pre-SN eruption event.

\section{Radiative-transfer simulation \label{sec:lcs}}

\subsection{Methodology}

A direct observational signature of the late-stage eruptions in CCSN progenitors is precursor electromagnetic transients \citep{2010MNRAS.405.2113D,2023ApJ...945..104T,2022ApJ...936..114M}. To model such transients, we perform radiative-transfer simulations using the multi-group radiative hydrodynamics code \texttt{STELLA} \citep{stella1,stella2,stella3}, available in the \texttt{MESA} software suite (r21.12.1; \citealt{MESA18}). As a widely-used code for modeling supernova light curves, \texttt{STELLA} solves the radiative transfer equation with a two-moment scheme that employs the variable Eddington factor method \citep{stella1}. It utilizes an EOS that includes ions, electrons, and positrons complemented with a Saha solver \citep{1994MNRAS.266..289B,2022A&A...668A.163B}, and accounts opacities from photoionization, free-free absorption, electron scattering, and line interactions \citep{1993ApJ...412..731E,1995all..book.....K,1995A&AS..109..125V,1996ADNDT..64....1V}. Radiation acceleration is computed using the first radiation moment (flux).

We extract the stellar structure from the eruption simulations when the shock reaches a mass coordinate of 0.2\,$M_\sun$ below the surface and use it as input for \texttt{STELLA}. The simulations preserve the original \texttt{SNEC} grid (cf. \S~\ref{ssec:method}) and employ 40 energy bins spanning 1-50000\,$\rm\AA$ for the spectral energy distribution (SED) of photons. We simulate only the 9.0\,$M_\sun$ model, as other models are expected to follow similar trends with $E_{\rm dep}$.

\begin{figure*}
    \centering
    \includegraphics[width=0.97\textwidth]{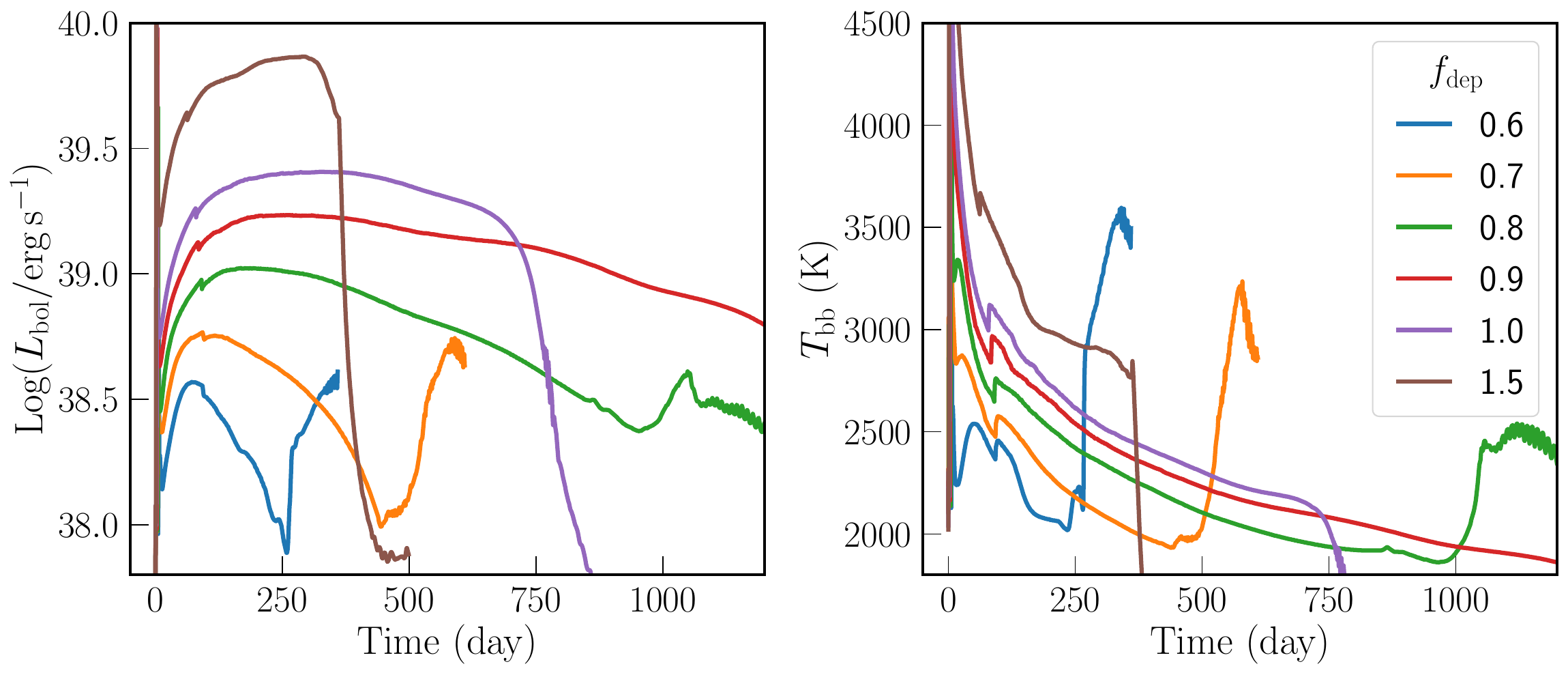}
    \caption{Bolometric light curves (left panel) and black-body temperature evolution (right panel) for the 9.0\,$M_\sun$ eruption models at selected values of $f_{\rm dep}$. Time zero is defined as the beginning of \texttt{STELLA} simulations.}
    \label{fig:lcs}
\end{figure*}

\subsection{Precursor signals}
Figure~\ref{fig:lcs} presents the bolometric light curves and black-body temperature ($T_{\rm bb}$) evolution of the precursor emission for selected values of $f_{\rm dep}$. The bolometric luminosity ($L_{\rm bol}$) is integrated from the SED in the outermost zone, and $T_{\rm bb}$ is derived from its black-body fit. Following a brief shock breakout pulse near time zero, all models exhibit a main emission phase lasting hundreds of days. The peak luminosities in this phase range from 10$^{38.5}$ to 10$^{40}$\,erg\,s$^{-1}$, approximately an order of magnitude fainter than the plateau of the observed least luminous SNe IIP \citep{2025arXiv250620068D}. As expected, a larger $f_{\rm dep}$ produces a brighter peak. As found in \cite{2010MNRAS.405.2113D}, the light-curve morphology evolves systematically with $f_{\rm dep}$. For small values (0.6-0.8), $L_{\rm bol}$ reaches the peak at $\sim$100-200\,days and then decays linearly in log space to $\sim10^{38}$\,erg\,s$^{-1}$. For an intermediate value ($f_{\rm dep}=0.9$), the decay is significantly slower and $L_{\rm bol}$ maintains $\sim10^{39}$\,erg\,s$^{-1}$ for more than 1000 days. At even larger values ($f_{\rm dep}\ge1.0$), the light curve resembles SNe IIP but with a fainter $L_{\rm bol}$ and a longer plateau ($\sim100$~days for typical SNe IIP). The time-integrated radiation energy ranges from $10^{46}$ to $10^{47}$\,erg, about 10\% of the energy gained by the H-rich envelope. Throughout the main emission phase, $T_{\rm bb}$ is in the range of 2000 to 3000\,K, cooler than the $\sim6000$\,K plateau of typical SNe IIP, and increases systematically with greater $f_{\rm dep}$.

As the radiation is primarily powered by the cooling of the H-rich ejecta, the dependence of light curves on $f_{\rm dep}$ directly reflects the influence of $f_{\rm dep}$ on the mass ejection. For a large $f_{\rm dep}$, the entire H-rich envelope is ejected and the light curves follow the scaling laws of SNe IIP (e.g., \citealt{1993ApJ...414..712P}). In this regime, the plateau luminosity increases and its duration decreases with higher explosion energy (represented here by $E_{\rm gain}$). For a small $f_{\rm dep}$, both the kinetic energy and the ejected mass are reduced. The lower ejected mass leads to a shorter photon diffusion time, producing a rapidly declining light curve. As the ejecta becomes optically thin and the bound material is revealed, the emission returns to its original RSG-like state. Lastly, the long-lasting emission for $f_{\rm dep}=0.9$ reflects the scenario in which a large fraction of the H-rich envelope is lost and expands over a very extended period (cf. Figure~\ref{fig:ej_vs_t}), sustaining the cooling emission. 

\begin{figure*}[t]
    \centering
    \includegraphics[width=0.97\textwidth]{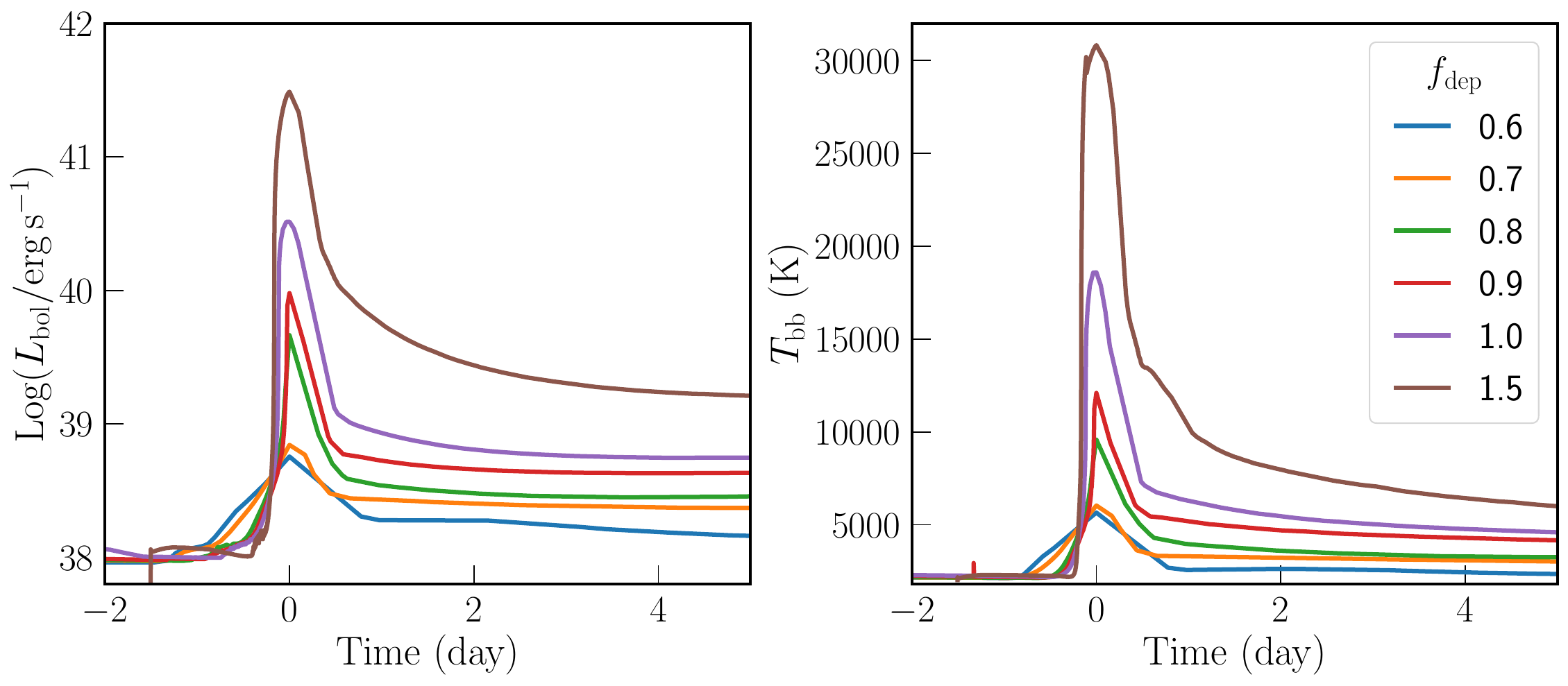}
    \caption{Bolometric light curves (left panel) and black-body temperature evolution (right panel) for the shock breakout signal of the 9.0\,$M_\sun$ eruption models at selected values of $f_{\rm dep}$. Time zero is defined as the epoch of peak bolometric luminosity.}
    \label{fig:sbo}
\end{figure*}

The brief shock breakout emission is highlighted in Figure~\ref{fig:sbo}, manifesting as a peak with a duration of $\sim1$ day. The peak luminosity increases with $f_{\rm dep}$, reaching values comparable to the plateau luminosities of low-luminosity SNe IIP for $f_{\rm dep}\ge1$. During this phase, $T_{\rm bb}$ peaks in the range of 5000-30000\,K, producing emission mainly in the optical to UV bands. Although this is brighter and resides more at short wavelengths compared to the main emission epoch, its short timescale renders it difficult to detect with current observing facilities, e.g., ZTF scans the whole northern sky every 2 nights. It can be a potential target for future high-cadence UV missions such as ULTRASAT \citep{ultrasat}.

\begin{figure*}[t]
    \centering
    \includegraphics[width=0.48\textwidth]{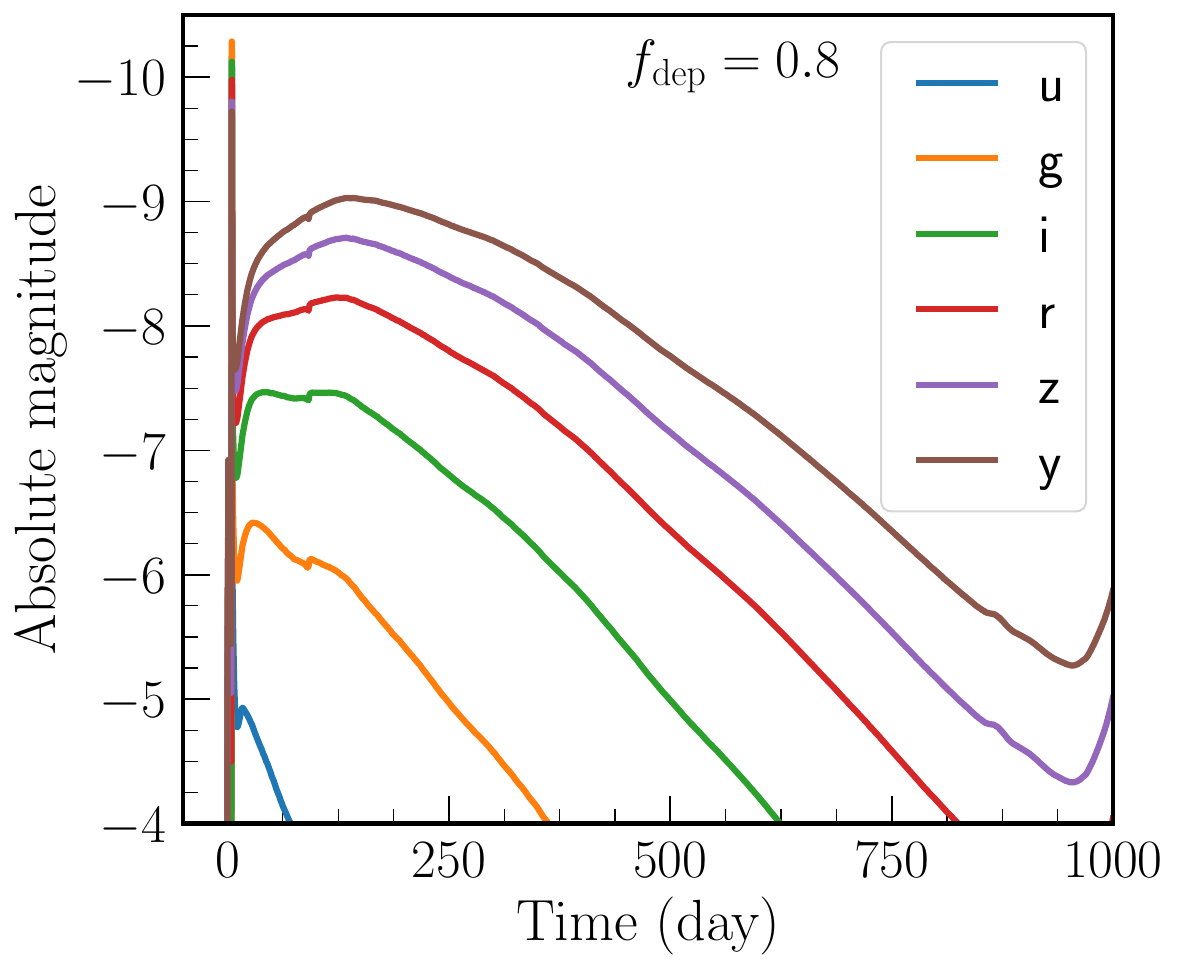}
    \includegraphics[width=0.48\textwidth]{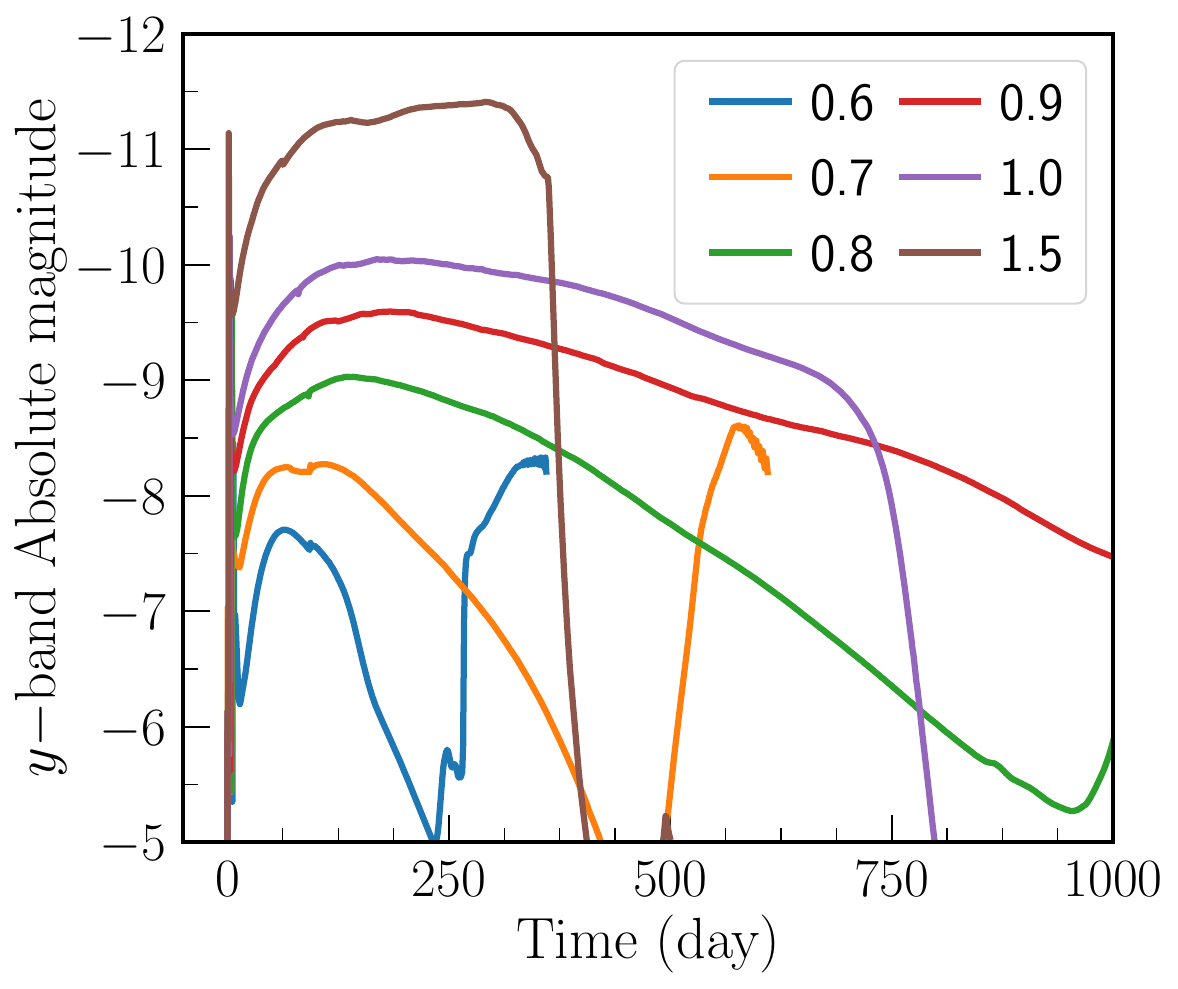}
    \caption{\textit{Left panel: } multi-band photometric light curves with the LSST filters for the 9.0\,$M_\sun$ model at $f_{\rm dep}=0.8$. \textit{Right panel: } photometric light curves in the LSST $y$-band for the 9.0\,$M_\sun$ model at selected values of $f_{\rm dep}$.}
    \label{fig:mlcs}
\end{figure*}

Finally, we generate multi-band photometric light curves for the stellar eruptions by convolving the SED with the filter transmissions of the Legacy Survey of Space and Time (LSST; \citealt{LSST}). The left panel shows a representative case of $f_{\rm dep}=0.8$. Due to the low $T_{\rm bb}$ ($\sim$2000-3000\,K), the absolute magnitude is higher for the more infrared bands, peaking at $M\sim$-9 for the LSST-$y$ band. Following the peak, the magnitudes decrease approximately linearly. The right panel illustrates the dependence of the $y$-band light curve on $f_{\rm dep}$, revealing a morphology that resembles the corresponding bolometric light curves (cf. Figure~\ref{fig:lcs}). Even the brightest case with $f_{\rm dep}=1.5$ in which the entire H-rich envelope is ejected, produces an eruption $\sim$2-3 magnitudes fainter than the SN precursors observed by ZTF ($M_r\lesssim-13$; \citealt{2021ApJ...907...99S}). Therefore, either a larger energy deposition or an additional radiation mechanism is required to account for the observed precursor luminosities. For example, increasing the deposited energy to $\sim10^{49}$\,erg ($f_{\rm dep}\approx5$) can produce an outburst with a magnitude of $M_r \approx -13$ and a duration of $\sim200$ days.

\section{Discussion \label{sec:discuss}}

Our study adopted a simplified and uniform approach by modeling the energy deposition as a single, sudden event at the base of the He layer. While this assumption is physically plausible for explosive nuclear burning, the reality may involve multiple flashes that require further dedicated stellar evolutionary modeling as \cite{2015ApJ...810...34W}. Previous works have provided hints on the impact of varying locations, durations, and number of deposition times \cite{2010MNRAS.405.2113D,2023ApJ...945..104T}. For instance, a longer deposition duration or a deeper location can yield outcomes equivalent to a smaller net energy input. \cite{2023ApJ...945..104T} demonstrated that double energy depositions can produce a significant enhancement of radiation when the second ejecta collides with the first. Ultimately, a comprehensive understanding of these eruptions will require future models that self-consistently incorporate the detailed properties of the underlying explosive nuclear flashes.

Apart from these uncertainties, we have chosen the deposited energy to be a fraction ($f_{\rm dep}$) of the total envelope binding energy $E_{\rm bind,tot}$. As noted in \S~\ref{ssec:method}, our choice of $f_{\rm dep}$ in the range of 0.5-1.0 results in the mass ejection from a non-negligible fraction ($\sim10^{-4}$) to the full envelope eruption (1.0). However, the energy released by explosive nuclear burning in the core is an intrinsic property, and it is not coupled to the envelope binding energy \citep{2015ApJ...810...34W}. Consequently, our exploration of the precursor light curves (cf.~Figure~\ref{fig:lcs}) is limited to the adopted $f_{\rm dep}$ range and does not reflect the full possible diversity of SN precursors. We note that with $f_{\rm dep}\approx 5$, the precursor luminosity can reach the lower limit of the observed precursor brightness by ZTF ($M_r\sim-13$; \citealt{2021ApJ...907...99S}). With even larger values of $f_{\rm dep}$, we expect the precursor light curve to resemble SN IIP, with its plateau luminosity and duration following the corresponding scaling relations \citep{1993ApJ...414..712P}. 

Although we performed radiative-transfer simulations, their primary purpose was to generate precursor light curves. Radiative cooling also influences the mass ejection process, slowing down the outflow and modifying its temperature profile. A complete model of the CSM and progenitor appearance should further account for subsequent processes such as dust formation \citep{2003ARA&A..41...15M,2005ApJ...634.1286M,2009A&A...498..127V}. To support such efforts, we provide publicly available profiles at selected epochs from both our \texttt{SNEC} and \texttt{STELLA} simulations.

Another important caveat of this study is the assumption of spherical symmetry in our hydrodynamic simulations. \citealt{2022ApJ...936...28T} has highlighted this limitation by performing a suite of 3D pre-SN eruption simulations with a single progenitor model of $M_{\rm ZAMS}=15\,M_\sun$. They demonstrated that multi-dimensional effects can lead to enhanced ejecta masses and an inhomogeneous structure in the CSM, with column densities varying by roughly an order of magnitude for different angles. However, conducting such 3D simulations is prohibitive for a wide parameter space, making them unfeasible for a systematic investigation of progenitor dependence and energy deposition parameters. A potential way forward is to develop calibrated 1D prescriptions that capture the essential effects of multi-dimensional dynamics informed by 3D simulations (e.g., \citealt{2016ApJ...821...76D}).

\section{Conclusions \label{sec:conclu}}

In this paper, we have investigated the consequences of sudden energy deposition at the base of the He layer in low-mass CCSN progenitors of \cite{2016ApJ...821...38S}, a scenario that mimics the energy release from late-stage explosive nuclear flashes \citep{1979BAAS...11Q.724W,2015ApJ...810...34W}.
Using non-radiative hydrodynamics simulations with \texttt{SNEC}, we traced the resulting outflow to determine the final ejecta mass (defined as shells exceeding the local escaping velocity), which typically saturates within 3 years post energy deposition. We established scaling relations between the ejecta mass and both the deposited energy and the energy gained by the H-rich envelope. For the latter, we found a power-law relation with an index of $\sim3.5$, similar to studies with polytropic stars (e.g., \citealt{2021MNRAS.501.4266L,2024ApJ...967...33C}). This relation shows limited scatter across the 9-10\,$M_\sun$ range considered, within a factor of $\sim2.6$. The shock passage flattens the bound H-rich envelope, a structural change that could affect the SN light-curve morphology and serves as an additional diagnostic for the eruption. We note that for deposited energies near the threshold of full envelope ejection, mass loss occurs over an extended period, making the determination of the final ejecta mass non-trivial.

Furthermore, we computed the associated precursor light curves for the low-mass progenitor considered (9\,$M_\sun$), using multi-group radiative hydrodynamics simulations with \texttt{STELLA}. These signals are faint, with luminosities of $\sim10^{39}$\,erg\,s$^{-1}$ sustained for hundreds of days. For low energy deposition, the light curve exhibits a linear decay after the main emission peak, with the star returning to its original RSG appearance once the ejecta becomes optically thin. In contrast, high energy deposition cases exhibit longer durations and plateau luminosities roughly an order of magnitude lower than the lowest luminosity SNe IIP \citep{2025arXiv250620068D} when the entire envelope is ejected. The precursor emission has a cool black-body temperature of $\sim$2000-3000\,K, causing it to peak in the infrared. Although the shock breakout signal can reach SN brightness, its brevity of $\sim$1 day makes it difficult to detect with current facilities. These findings are consistent with results from more massive progenitors reported in \cite{2010MNRAS.405.2113D,2023ApJ...945..104T}.

Our simulations are designed to follow the process of mass ejection and its associated radiation properties. However, the energy initially deposited is significantly consumed by the material that remains bound. The energy carried away by the ejecta and its radiated energy is a fraction of the deposited energy. For cases of weak energy deposition, only a tiny fraction is radiated promptly. For example, with $f_{\rm dep}=0.6$, the radiated energy is less than $10^{46}$\,erg, compared to the deposited energy of $\sim10^{48}$\,erg. The remaining energy may be stored in the bound envelope as thermal energy and can later be radiated away as the ejecta becomes optically thin. To fully understand the impact of inner energy deposition on pre-SN emission, it is essential to model the secular evolution of the perturbed bound envelope. Considering the long timescale, this is most feasibly achieved with quasi-static radiation transfer. 

\section*{Data availability}
The numerical results of this work, including the profiles of the evolved envelope and circumstellar material, and light curves of the eruptions, are publicly available at the Science Data Bank doi: \href{doi.org/10.57760/sciencedb.31295}{10.57760/sciencedb.31295}. 

\begin{acknowledgments}
We thank the anonymous referee for their constructive comments. We thank Dong Lai and Evan O'Connor for their stimulating conversations. This work is supported by the National Natural Science Foundation of China (NSFC, grant Nos. 12288102, 12125303, 12090040/3, 12393811, 12473031), the Strategic Priority Research
Program of the Chinese Academy of Sciences (grant Nos. XDB1160301, XDB1160300, XDB1160000), the National Key R\&D Program of China (grant No. 2021YFA1600401/2021YFA1600403), Yunnan Key Laboratory of Supernova Research (No. 202505AV340004), the Yunnan Revitalization Talent Support Program—Science \& Technology Champion Project (No.202305AB35003), the Yunnan Revitalization Talent Support Program--Young Talent project, and the Yunnan Fundamental Research Projects (grant NOs. 202501AS070078, 202401BC070007). The authors gratefully acknowledge the “PHOENIX Supercomputing Platform” jointly operated by the Binary Population Synthesis Group and the Stellar Astrophysics Group at Yunnan Observatories, CAS. 
H. L. was supported by the National Natural Science Foundation of China (NSFC grants No. 12403061) and the innovative project of ‘Caiyun Postdoctoral Project’ of Yunnan Province.
\end{acknowledgments}

\vspace{5mm}

\software{ \texttt{SNEC} \citep{2015ApJ...814...63M}; \texttt{STELLA} \citep{stella1,stella2,stella3,MESA18}; \texttt{Astropy} \citep{astropy}; \texttt{Numpy} \citep{numpy};  \texttt{Matplotlib} \citep{matplotlib}}

\bibliography{presn}{}

@ARTICLE{ZTF,
       author = {{Graham}, Matthew J. and {Kulkarni}, S.~R. and {Bellm}, Eric C. and {Adams}, Scott M. and {Barbarino}, Cristina and {Blagorodnova}, Nadejda and {Bodewits}, Dennis and {Bolin}, Bryce and {Brady}, Patrick R. and {Cenko}, S. Bradley and {Chang}, Chan-Kao and {Coughlin}, Michael W. and {De}, Kishalay and {Eadie}, Gwendolyn and {Farnham}, Tony L. and {Feindt}, Ulrich and {Franckowiak}, Anna and {Fremling}, Christoffer and {Gezari}, Suvi and {Ghosh}, Shaon and {Goldstein}, Daniel A. and {Golkhou}, V. Zach and {Goobar}, Ariel and {Ho}, Anna Y.~Q. and {Huppenkothen}, Daniela and {Ivezi{\'c}}, {\v{Z}}eljko and {Jones}, R. Lynne and {Juric}, Mario and {Kaplan}, David L. and {Kasliwal}, Mansi M. and {Kelley}, Michael S.~P. and {Kupfer}, Thomas and {Lee}, Chien-De and {Lin}, Hsing Wen and {Lunnan}, Ragnhild and {Mahabal}, Ashish A. and {Miller}, Adam A. and {Ngeow}, Chow-Choong and {Nugent}, Peter and {Ofek}, Eran O. and {Prince}, Thomas A. and {Rauch}, Ludwig and {van Roestel}, Jan and {Schulze}, Steve and {Singer}, Leo P. and {Sollerman}, Jesper and {Taddia}, Francesco and {Yan}, Lin and {Ye}, Quan-Zhi and {Yu}, Po-Chieh and {Barlow}, Tom and {Bauer}, James and {Beck}, Ron and {Belicki}, Justin and {Biswas}, Rahul and {Brinnel}, Valery and {Brooke}, Tim and {Bue}, Brian and {Bulla}, Mattia and {Burruss}, Rick and {Connolly}, Andrew and {Cromer}, John and {Cunningham}, Virginia and {Dekany}, Richard and {Delacroix}, Alex and {Desai}, Vandana and {Duev}, Dmitry A. and {Feeney}, Michael and {Flynn}, David and {Frederick}, Sara and {Gal-Yam}, Avishay and {Giomi}, Matteo and {Groom}, Steven and {Hacopians}, Eugean and {Hale}, David and {Helou}, George and {Henning}, John and {Hover}, David and {Hillenbrand}, Lynne A. and {Howell}, Justin and {Hung}, Tiara and {Imel}, David and {Ip}, Wing-Huen and {Jackson}, Edward and {Kaspi}, Shai and {Kaye}, Stephen and {Kowalski}, Marek and {Kramer}, Emily and {Kuhn}, Michael and {Landry}, Walter and {Laher}, Russ R. and {Mao}, Peter and {Masci}, Frank J. and {Monkewitz}, Serge and {Murphy}, Patrick and {Nordin}, Jakob and {Patterson}, Maria T. and {Penprase}, Bryan and {Porter}, Michael and {Rebbapragada}, Umaa and {Reiley}, Dan and {Riddle}, Reed and {Rigault}, Mickael and {Rodriguez}, Hector and {Rusholme}, Ben and {van Santen}, Jakob and {Shupe}, David L. and {Smith}, Roger M. and {Soumagnac}, Maayane T. and {Stein}, Robert and {Surace}, Jason and {Szkody}, Paula and {Terek}, Scott and {Van Sistine}, Angela and {van Velzen}, Sjoert and {Vestrand}, W. Thomas and {Walters}, Richard and {Ward}, Charlotte and {Zhang}, Chaoran and {Zolkower}, Jeffry},
        title = "{The Zwicky Transient Facility: Science Objectives}",
      journal = {\pasp},
         year = 2019,
        month = jul,
       volume = {131},
       number = {1001},
        pages = {078001},
          doi = {10.1088/1538-3873/ab006c},
archivePrefix = {arXiv},
       eprint = {1902.01945},
 primaryClass = {astro-ph.IM},
       adsurl = {https://ui.adsabs.harvard.edu/abs/2019PASP..131g8001G}
}

@ARTICLE{WFST,
       author = {{Wang}, Tinggui and {Liu}, Guilin and {Cai}, Zhenyi and {Geng}, Jinjun and {Fang}, Min and {He}, Haoning and {Jiang}, Ji-an and {Jiang}, Ning and {Kong}, Xu and {Li}, Bin and {Li}, Ye and {Luo}, Wentao and {Pan}, Zhizheng and {Wu}, Xuefeng and {Yang}, Ji and {Yu}, Jiming and {Zheng}, Xianzhong and {Zhu}, Qingfeng and {Cai}, Yi-Fu and {Chen}, Yuanyuan and {Chen}, Zhiwei and {Dai}, Zigao and {Fan}, Lulu and {Fan}, Yizhong and {Fang}, Wenjuan and {He}, Zhicheng and {Hu}, Lei and {Hu}, Maokai and {Jin}, Zhiping and {Jiang}, Zhibo and {Li}, Guoliang and {Li}, Fan and {Li}, Xuzhi and {Liang}, Runduo and {Lin}, Zheyu and {Liu}, Qingzhong and {Liu}, Wenhao and {Liu}, Zhengyan and {Liu}, Wei and {Liu}, Yao and {Lou}, Zheng and {Qu}, Han and {Sheng}, Zhenfeng and {Shi}, Jianchun and {Shu}, Yiping and {Su}, Zhenbo and {Sun}, Tianrui and {Wang}, Hongchi and {Wang}, Huiyuan and {Wang}, Jian and {Wang}, Junxian and {Wei}, Daming and {Wei}, Junjie and {Xue}, Yongquan and {Yan}, Jingzhi and {Yang}, Chao and {Yuan}, Ye and {Yuan}, Yefei and {Zhang}, Hongxin and {Zhang}, Miaomiao and {Zhao}, Haibin and {Zhao}, Wen},
        title = "{Science with the 2.5-meter Wide Field Survey Telescope (WFST)}",
      journal = {Science China Physics, Mechanics, and Astronomy},
         year = 2023,
        month = oct,
       volume = {66},
       number = {10},
          eid = {109512},
        pages = {109512},
          doi = {10.1007/s11433-023-2197-5},
archivePrefix = {arXiv},
       eprint = {2306.07590},
 primaryClass = {astro-ph.IM},
       adsurl = {https://ui.adsabs.harvard.edu/abs/2023SCPMA..6609512W}
}

@ARTICLE{LSST,
       author = {{Ivezi{\'c}}, {\v{Z}}eljko and {Kahn}, Steven M. and {Tyson}, J. Anthony and {Abel}, Bob and {Acosta}, Emily and {Allsman}, Robyn and {Alonso}, David and {AlSayyad}, Yusra and {Anderson}, Scott F. and {Andrew}, John and {Angel}, James Roger P. and {Angeli}, George Z. and {Ansari}, Reza and {Antilogus}, Pierre and {Araujo}, Constanza and {Armstrong}, Robert and {Arndt}, Kirk T. and {Astier}, Pierre and {Aubourg}, {\'E}ric and {Auza}, Nicole and {Axelrod}, Tim S. and {Bard}, Deborah J. and {Barr}, Jeff D. and {Barrau}, Aurelian and {Bartlett}, James G. and {Bauer}, Amanda E. and {Bauman}, Brian J. and {Baumont}, Sylvain and {Bechtol}, Ellen and {Bechtol}, Keith and {Becker}, Andrew C. and {Becla}, Jacek and {Beldica}, Cristina and {Bellavia}, Steve and {Bianco}, Federica B. and {Biswas}, Rahul and {Blanc}, Guillaume and {Blazek}, Jonathan and {Blandford}, Roger D. and {Bloom}, Josh S. and {Bogart}, Joanne and {Bond}, Tim W. and {Booth}, Michael T. and {Borgland}, Anders W. and {Borne}, Kirk and {Bosch}, James F. and {Boutigny}, Dominique and {Brackett}, Craig A. and {Bradshaw}, Andrew and {Brandt}, William Nielsen and {Brown}, Michael E. and {Bullock}, James S. and {Burchat}, Patricia and {Burke}, David L. and {Cagnoli}, Gianpietro and {Calabrese}, Daniel and {Callahan}, Shawn and {Callen}, Alice L. and {Carlin}, Jeffrey L. and {Carlson}, Erin L. and {Chandrasekharan}, Srinivasan and {Charles-Emerson}, Glenaver and {Chesley}, Steve and {Cheu}, Elliott C. and {Chiang}, Hsin-Fang and {Chiang}, James and {Chirino}, Carol and {Chow}, Derek and {Ciardi}, David R. and {Claver}, Charles F. and {Cohen-Tanugi}, Johann and {Cockrum}, Joseph J. and {Coles}, Rebecca and {Connolly}, Andrew J. and {Cook}, Kem H. and {Cooray}, Asantha and {Covey}, Kevin R. and {Cribbs}, Chris and {Cui}, Wei and {Cutri}, Roc and {Daly}, Philip N. and {Daniel}, Scott F. and {Daruich}, Felipe and {Daubard}, Guillaume and {Daues}, Greg and {Dawson}, William and {Delgado}, Francisco and {Dellapenna}, Alfred and {de Peyster}, Robert and {de Val-Borro}, Miguel and {Digel}, Seth W. and {Doherty}, Peter and {Dubois}, Richard and {Dubois-Felsmann}, Gregory P. and {Durech}, Josef and {Economou}, Frossie and {Eifler}, Tim and {Eracleous}, Michael and {Emmons}, Benjamin L. and {Fausti Neto}, Angelo and {Ferguson}, Henry and {Figueroa}, Enrique and {Fisher-Levine}, Merlin and {Focke}, Warren and {Foss}, Michael D. and {Frank}, James and {Freemon}, Michael D. and {Gangler}, Emmanuel and {Gawiser}, Eric and {Geary}, John C. and {Gee}, Perry and {Geha}, Marla and {Gessner}, Charles J.~B. and {Gibson}, Robert R. and {Gilmore}, D. Kirk and {Glanzman}, Thomas and {Glick}, William and {Goldina}, Tatiana and {Goldstein}, Daniel A. and {Goodenow}, Iain and {Graham}, Melissa L. and {Gressler}, William J. and {Gris}, Philippe and {Guy}, Leanne P. and {Guyonnet}, Augustin and {Haller}, Gunther and {Harris}, Ron and {Hascall}, Patrick A. and {Haupt}, Justine and {Hernandez}, Fabio and {Herrmann}, Sven and {Hileman}, Edward and {Hoblitt}, Joshua and {Hodgson}, John A. and {Hogan}, Craig and {Howard}, James D. and {Huang}, Dajun and {Huffer}, Michael E. and {Ingraham}, Patrick and {Innes}, Walter R. and {Jacoby}, Suzanne H. and {Jain}, Bhuvnesh and {Jammes}, Fabrice and {Jee}, M. James and {Jenness}, Tim and {Jernigan}, Garrett and {Jevremovi{\'c}}, Darko and {Johns}, Kenneth and {Johnson}, Anthony S. and {Johnson}, Margaret W.~G. and {Jones}, R. Lynne and {Juramy-Gilles}, Claire and {Juri{\'c}}, Mario and {Kalirai}, Jason S. and {Kallivayalil}, Nitya J. and {Kalmbach}, Bryce and {Kantor}, Jeffrey P. and {Karst}, Pierre and {Kasliwal}, Mansi M. and {Kelly}, Heather and {Kessler}, Richard and {Kinnison}, Veronica and {Kirkby}, David and {Knox}, Lloyd and {Kotov}, Ivan V. and {Krabbendam}, Victor L. and {Krughoff}, K. Simon and {Kub{\'a}nek}, Petr and {Kuczewski}, John and {Kulkarni}, Shri and {Ku}, John and {Kurita}, Nadine R. and {Lage}, Craig S. and {Lambert}, Ron and {Lange}, Travis and {Langton}, J. Brian and {Le Guillou}, Laurent and {Levine}, Deborah and {Liang}, Ming and {Lim}, Kian-Tat and {Lintott}, Chris J. and {Long}, Kevin E. and {Lopez}, Margaux and {Lotz}, Paul J. and {Lupton}, Robert H. and {Lust}, Nate B. and {MacArthur}, Lauren A. and {Mahabal}, Ashish and {Mandelbaum}, Rachel and {Markiewicz}, Thomas W. and {Marsh}, Darren S. and {Marshall}, Philip J. and {Marshall}, Stuart and {May}, Morgan and {McKercher}, Robert and {McQueen}, Michelle and {Meyers}, Joshua and {Migliore}, Myriam and {Miller}, Michelle and {Mills}, David J.},
        title = "{LSST: From Science Drivers to Reference Design and Anticipated Data Products}",
      journal = {\apj},
         year = 2019,
        month = mar,
       volume = {873},
       number = {2},
          eid = {111},
        pages = {111},
          doi = {10.3847/1538-4357/ab042c},
archivePrefix = {arXiv},
       eprint = {0805.2366},
 primaryClass = {astro-ph},
       adsurl = {https://ui.adsabs.harvard.edu/abs/2019ApJ...873..111I}
}

@INPROCEEDINGS{MEPHISTO,
       author = {{Yuan}, Xiangyan and {Li}, Zhengyang and {Liu}, Xiaowei and {Niu}, Dongsheng and {Lu}, Qishui and {Jiang}, Fanghua and {Wang}, Yuefei and {Li}, Xiaoyan and {Liang}, YongJun and {Wang}, Hai and {Zhang}, Chao and {Wang}, Jinfeng and {Li}, Bo and {Tian}, Jie and {Lu}, Haiping and {Chen}, Bingqiu and {Huang}, Yang and {Liu}, Xiangkun and {Yao}, Zhengqiu and {Cui}, Xiangqun and {Li}, Guoping},
        title = "{Development of the Multi-channel Photometric Survey telescope}",
    booktitle = {Ground-based and Airborne Telescopes VIII},
         year = 2020,
       editor = {{Marshall}, Heather K. and {Spyromilio}, Jason and {Usuda}, Tomonori},
       series = {Society of Photo-Optical Instrumentation Engineers (SPIE) Conference Series},
       volume = {11445},
        month = dec,
          eid = {114457M},
        pages = {114457M},
          doi = {10.1117/12.2562334},
       adsurl = {https://ui.adsabs.harvard.edu/abs/2020SPIE11445E..7MY}
}

@ARTICLE{ATLAS,
       author = {{Tonry}, J.~L. and {Denneau}, L. and {Heinze}, A.~N. and {Stalder}, B. and {Smith}, K.~W. and {Smartt}, S.~J. and {Stubbs}, C.~W. and {Weiland}, H.~J. and {Rest}, A.},
        title = "{ATLAS: A High-cadence All-sky Survey System}",
      journal = {\pasp},
         year = 2018,
        month = jun,
       volume = {130},
       number = {988},
        pages = {064505},
          doi = {10.1088/1538-3873/aabadf},
archivePrefix = {arXiv},
       eprint = {1802.00879},
 primaryClass = {astro-ph.IM},
       adsurl = {https://ui.adsabs.harvard.edu/abs/2018PASP..130f4505T}
}

@ARTICLE{PANSTARS,
       author = {{Chambers}, K.~C. and {Magnier}, E.~A. and {Metcalfe}, N. and {Flewelling}, H.~A. and {Huber}, M.~E. and {Waters}, C.~Z. and {Denneau}, L. and {Draper}, P.~W. and {Farrow}, D. and {Finkbeiner}, D.~P. and {Holmberg}, C. and {Koppenhoefer}, J. and {Price}, P.~A. and {Rest}, A. and {Saglia}, R.~P. and {Schlafly}, E.~F. and {Smartt}, S.~J. and {Sweeney}, W. and {Wainscoat}, R.~J. and {Burgett}, W.~S. and {Chastel}, S. and {Grav}, T. and {Heasley}, J.~N. and {Hodapp}, K.~W. and {Jedicke}, R. and {Kaiser}, N. and {Kudritzki}, R. -P. and {Luppino}, G.~A. and {Lupton}, R.~H. and {Monet}, D.~G. and {Morgan}, J.~S. and {Onaka}, P.~M. and {Shiao}, B. and {Stubbs}, C.~W. and {Tonry}, J.~L. and {White}, R. and {Ba{\~n}ados}, E. and {Bell}, E.~F. and {Bender}, R. and {Bernard}, E.~J. and {Boegner}, M. and {Boffi}, F. and {Botticella}, M.~T. and {Calamida}, A. and {Casertano}, S. and {Chen}, W. -P. and {Chen}, X. and {Cole}, S. and {Deacon}, N. and {Frenk}, C. and {Fitzsimmons}, A. and {Gezari}, S. and {Gibbs}, V. and {Goessl}, C. and {Goggia}, T. and {Gourgue}, R. and {Goldman}, B. and {Grant}, P. and {Grebel}, E.~K. and {Hambly}, N.~C. and {Hasinger}, G. and {Heavens}, A.~F. and {Heckman}, T.~M. and {Henderson}, R. and {Henning}, T. and {Holman}, M. and {Hopp}, U. and {Ip}, W. -H. and {Isani}, S. and {Jackson}, M. and {Keyes}, C.~D. and {Koekemoer}, A.~M. and {Kotak}, R. and {Le}, D. and {Liska}, D. and {Long}, K.~S. and {Lucey}, J.~R. and {Liu}, M. and {Martin}, N.~F. and {Masci}, G. and {McLean}, B. and {Mindel}, E. and {Misra}, P. and {Morganson}, E. and {Murphy}, D.~N.~A. and {Obaika}, A. and {Narayan}, G. and {Nieto-Santisteban}, M.~A. and {Norberg}, P. and {Peacock}, J.~A. and {Pier}, E.~A. and {Postman}, M. and {Primak}, N. and {Rae}, C. and {Rai}, A. and {Riess}, A. and {Riffeser}, A. and {Rix}, H.~W. and {R{\"o}ser}, S. and {Russel}, R. and {Rutz}, L. and {Schilbach}, E. and {Schultz}, A.~S.~B. and {Scolnic}, D. and {Strolger}, L. and {Szalay}, A. and {Seitz}, S. and {Small}, E. and {Smith}, K.~W. and {Soderblom}, D.~R. and {Taylor}, P. and {Thomson}, R. and {Taylor}, A.~N. and {Thakar}, A.~R. and {Thiel}, J. and {Thilker}, D. and {Unger}, D. and {Urata}, Y. and {Valenti}, J. and {Wagner}, J. and {Walder}, T. and {Walter}, F. and {Watters}, S.~P. and {Werner}, S. and {Wood-Vasey}, W.~M. and {Wyse}, R.},
        title = "{The Pan-STARRS1 Surveys}",
      journal = {arXiv e-prints},
         year = 2016,
        month = dec,
          eid = {arXiv:1612.05560},
        pages = {arXiv:1612.05560},
          doi = {10.48550/arXiv.1612.05560},
archivePrefix = {arXiv},
       eprint = {1612.05560},
 primaryClass = {astro-ph.IM},
       adsurl = {https://ui.adsabs.harvard.edu/abs/2016arXiv161205560C}
}

@INPROCEEDINGS{ASASSN,
       author = {{Shappee}, Benjamin and {Prieto}, J. and {Stanek}, K.~Z. and {Kochanek}, C.~S. and {Holoien}, T. and {Jencson}, J. and {Basu}, U. and {Beacom}, J.~F. and {Szczygiel}, D. and {Pojmanski}, G. and {Brimacombe}, J. and {Dubberley}, M. and {Elphick}, M. and {Foale}, S. and {Hawkins}, E. and {Mullins}, D. and {Rosing}, W. and {Ross}, R. and {Walker}, Z.},
        title = "{All Sky Automated Survey for SuperNovae (ASAS-SN or ``Assassin'')}",
    booktitle = {American Astronomical Society Meeting Abstracts \#223},
         year = 2014,
       series = {American Astronomical Society Meeting Abstracts},
       volume = {223},
        month = jan,
          eid = {236.03},
        pages = {236.03},
       adsurl = {https://ui.adsabs.harvard.edu/abs/2014AAS...22323603S}
}

@ARTICLE{2023ApJ...952..119B,
       author = {{Bruch}, Rachel J. and {Gal-Yam}, Avishay and {Yaron}, Ofer and {Chen}, Ping and {Strotjohann}, Nora L. and {Irani}, Ido and {Zimmerman}, Erez and {Schulze}, Steve and {Yang}, Yi and {Kim}, Young-Lo and {Bulla}, Mattia and {Sollerman}, Jesper and {Rigault}, Mickael and {Ofek}, Eran and {Soumagnac}, Maayane and {Masci}, Frank J. and {Fremling}, Christoffer and {Perley}, Daniel and {Nordin}, Jakob and {Cenko}, S. Bradley and {Ho}, Anna Y.~Q. and {Adams}, S. and {Adreoni}, Igor and {Bellm}, Eric C. and {Blagorodnova}, Nadia and {Burdge}, Kevin and {De}, Kishalay and {Dekany}, Richard G. and {Dhawan}, Suhail and {Drake}, Andrew J. and {Duev}, Dmitry A. and {Graham}, Matthew and {Graham}, Melissa L. and {Jencson}, Jacob and {Karamehmetoglu}, Emir and {Kasliwal}, Mansi M. and {Kulkarni}, Shrinivas and {Miller}, A.~A. and {Neill}, James D. and {Prince}, Thomas A. and {Riddle}, Reed and {Rusholme}, Benjamin and {Sharma}, Y. and {Smith}, Roger and {Sravan}, Niharika and {Taggart}, Kirsty and {Walters}, Richard and {Yan}, Lin},
        title = "{The Prevalence and Influence of Circumstellar Material around Hydrogen-rich Supernova Progenitors}",
      journal = {\apj},
         year = 2023,
        month = aug,
       volume = {952},
       number = {2},
          eid = {119},
        pages = {119},
          doi = {10.3847/1538-4357/acd8be},
archivePrefix = {arXiv},
       eprint = {2212.03313},
 primaryClass = {astro-ph.HE},
       adsurl = {https://ui.adsabs.harvard.edu/abs/2023ApJ...952..119B}
}

@ARTICLE{2021ApJ...912...46B,
       author = {{Bruch}, Rachel J. and {Gal-Yam}, Avishay and {Schulze}, Steve and {Yaron}, Ofer and {Yang}, Yi and {Soumagnac}, Maayane and {Rigault}, Mickael and {Strotjohann}, Nora L. and {Ofek}, Eran and {Sollerman}, Jesper and {Masci}, Frank J. and {Barbarino}, Cristina and {Ho}, Anna Y.~Q. and {Fremling}, Christoffer and {Perley}, Daniel and {Nordin}, Jakob and {Cenko}, S. Bradley and {Adams}, S. and {Adreoni}, Igor and {Bellm}, Eric C. and {Blagorodnova}, Nadia and {Bulla}, Mattia and {Burdge}, Kevin and {De}, Kishalay and {Dhawan}, Suhail and {Drake}, Andrew J. and {Duev}, Dmitry A. and {Dugas}, Alison and {Graham}, Matthew and {Graham}, Melissa L. and {Irani}, Ido and {Jencson}, Jacob and {Karamehmetoglu}, Emir and {Kasliwal}, Mansi and {Kim}, Young-Lo and {Kulkarni}, Shrinivas and {Kupfer}, Thomas and {Liang}, Jingyi and {Mahabal}, Ashish and {Miller}, A.~A. and {Prince}, Thomas A. and {Riddle}, Reed and {Sharma}, Y. and {Smith}, Roger and {Taddia}, Francesco and {Taggart}, Kirsty and {Walters}, Richard and {Yan}, Lin},
        title = "{A Large Fraction of Hydrogen-rich Supernova Progenitors Experience Elevated Mass Loss Shortly Prior to Explosion}",
      journal = {\apj},
         year = 2021,
        month = may,
       volume = {912},
       number = {1},
          eid = {46},
        pages = {46},
          doi = {10.3847/1538-4357/abef05},
archivePrefix = {arXiv},
       eprint = {2008.09986},
 primaryClass = {astro-ph.HE},
       adsurl = {https://ui.adsabs.harvard.edu/abs/2021ApJ...912...46B}
}

@ARTICLE{2022MNRAS.517.1483D,
       author = {{Davies}, Ben and {Plez}, Bertrand and {Petrault}, Mike},
        title = "{Explosion imminent: the appearance of red supergiants at the point of core-collapse}",
      journal = {\mnras},
         year = 2022,
        month = nov,
       volume = {517},
       number = {1},
        pages = {1483-1490},
          doi = {10.1093/mnras/stac2427},
archivePrefix = {arXiv},
       eprint = {2208.10883},
 primaryClass = {astro-ph.SR},
       adsurl = {https://ui.adsabs.harvard.edu/abs/2022MNRAS.517.1483D}
}

@INCOLLECTION{2017hsn..book..403S,
       author = {{Smith}, Nathan},
        title = "{Interacting Supernovae: Types IIn and Ibn}",
    booktitle = {Handbook of Supernovae},
         year = 2017,
       editor = {{Alsabti}, Athem W. and {Murdin}, Paul},
        pages = {403},
          doi = {10.1007/978-3-319-21846-5_38},
       adsurl = {https://ui.adsabs.harvard.edu/abs/2017hsn..book..403S}
}

@ARTICLE{2015ApJ...814...63M,
       author = {{Morozova}, Viktoriya and {Piro}, Anthony L. and {Renzo}, Mathieu and {Ott}, Christian D. and {Clausen}, Drew and {Couch}, Sean M. and {Ellis}, Justin and {Roberts}, Luke F.},
        title = "{Light Curves of Core-collapse Supernovae with Substantial Mass Loss Using the New Open-source SuperNova Explosion Code (SNEC)}",
      journal = {\apj},
         year = 2015,
        month = nov,
       volume = {814},
       number = {1},
          eid = {63},
        pages = {63},
          doi = {10.1088/0004-637X/814/1/63},
archivePrefix = {arXiv},
       eprint = {1505.06746},
 primaryClass = {astro-ph.HE},
       adsurl = {https://ui.adsabs.harvard.edu/abs/2015ApJ...814...63M}
}

@ARTICLE{stella3,
       author = {{Blinnikov}, S.~I. and {R{\"o}pke}, F.~K. and {Sorokina}, E.~I. and {Gieseler}, M. and {Reinecke}, M. and {Travaglio}, C. and {Hillebrandt}, W. and {Stritzinger}, M.},
        title = "{Theoretical light curves for deflagration models of type Ia supernova}",
      journal = {\aap},
         year = 2006,
        month = jul,
       volume = {453},
       number = {1},
        pages = {229-240},
          doi = {10.1051/0004-6361:20054594},
archivePrefix = {arXiv},
       eprint = {astro-ph/0603036},
 primaryClass = {astro-ph},
       adsurl = {https://ui.adsabs.harvard.edu/abs/2006A&A...453..229B}
}

@ARTICLE{stella2,
       author = {{Blinnikov}, Sergei and {Lundqvist}, Peter and {Bartunov}, Oleg and {Nomoto}, Ken'ichi and {Iwamoto}, Koichi},
        title = "{Radiation Hydrodynamics of SN 1987A. I. Global Analysis of the Light Curve for the First 4 Months}",
      journal = {\apj},
         year = 2000,
        month = apr,
       volume = {532},
       number = {2},
        pages = {1132-1149},
          doi = {10.1086/308588},
archivePrefix = {arXiv},
       eprint = {astro-ph/9911205},
 primaryClass = {astro-ph},
       adsurl = {https://ui.adsabs.harvard.edu/abs/2000ApJ...532.1132B}
}

@ARTICLE{stella1,
       author = {{Blinnikov}, S.~I. and {Eastman}, R. and {Bartunov}, O.~S. and {Popolitov}, V.~A. and {Woosley}, S.~E.},
        title = "{A Comparative Modeling of Supernova 1993J}",
      journal = {\apj},
         year = 1998,
        month = mar,
       volume = {496},
       number = {1},
        pages = {454-472},
          doi = {10.1086/305375},
archivePrefix = {arXiv},
       eprint = {astro-ph/9711055},
 primaryClass = {astro-ph},
       adsurl = {https://ui.adsabs.harvard.edu/abs/1998ApJ...496..454B}
}

@ARTICLE{MESA18,
       author = {{Paxton}, Bill and {Schwab}, Josiah and {Bauer}, Evan B. and {Bildsten}, Lars and {Blinnikov}, Sergei and {Duffell}, Paul and {Farmer}, R. and {Goldberg}, Jared A. and {Marchant}, Pablo and {Sorokina}, Elena and {Thoul}, Anne and {Townsend}, Richard H.~D. and {Timmes}, F.~X.},
        title = "{Modules for Experiments in Stellar Astrophysics (MESA): Convective Boundaries, Element Diffusion, and Massive Star Explosions}",
      journal = {\apjs},
         year = 2018,
        month = feb,
       volume = {234},
       number = {2},
          eid = {34},
        pages = {34},
          doi = {10.3847/1538-4365/aaa5a8},
archivePrefix = {arXiv},
       eprint = {1710.08424},
 primaryClass = {astro-ph.SR},
       adsurl = {https://ui.adsabs.harvard.edu/abs/2018ApJS..234...34P}
}

@article{numpy,
	title="Array programming with NumPy.",
	author="Charles R. {Harris} and K. Jarrod {Millman} and Stéfan J. van der {Walt} and Ralf {Gommers} and Pauli {Virtanen} and David {Cournapeau} and Eric {Wieser} and Julian {Taylor} and Sebastian {Berg} and Nathaniel J. {Smith} and Robert {Kern} and Matti {Picus} and Stephan {Hoyer} and Marten H. van {Kerkwijk} and Matthew {Brett} and Allan {Haldane} and Jaime Fernández del {Río} and Mark {Wiebe} and Pearu {Peterson} and Pierre {Gérard-Marchant} and Kevin {Sheppard} and Tyler {Reddy} and Warren {Weckesser} and Hameer {Abbasi} and Christoph {Gohlke} and Travis E. {Oliphant}",
	journal="Nature",
	volume="585",
	number="7825",
	pages="357--362",
	notes="Sourced from Microsoft Academic - https://academic.microsoft.com/paper/3099878876",
	year="2020"
}

@Article{matplotlib,
  Author    = {Hunter, J. D.},
  Title     = {Matplotlib: A 2D graphics environment},
  Journal   = {Computing in Science \& Engineering},
  Volume    = {9},
  Number    = {3},
  Pages     = {90--95},
  publisher = {IEEE COMPUTER SOC},
  doi       = {10.1109/MCSE.2007.55},
  year      = 2007
}

@ARTICLE{astropy,
       author = {{Astropy Collaboration} and {Price-Whelan}, Adrian M. and {Lim}, Pey Lian and {Earl}, Nicholas and {Starkman}, Nathaniel and {Bradley}, Larry and {Shupe}, David L. and {Patil}, Aarya A. and {Corrales}, Lia and {Brasseur}, C.~E. and {N{\"o}the}, Maximilian and {Donath}, Axel and {Tollerud}, Erik and {Morris}, Brett M. and {Ginsburg}, Adam and {Vaher}, Eero and {Weaver}, Benjamin A. and {Tocknell}, James and {Jamieson}, William and {van Kerkwijk}, Marten H. and {Robitaille}, Thomas P. and {Merry}, Bruce and {Bachetti}, Matteo and {G{\"u}nther}, H. Moritz and {Aldcroft}, Thomas L. and {Alvarado-Montes}, Jaime A. and {Archibald}, Anne M. and {B{\'o}di}, Attila and {Bapat}, Shreyas and {Barentsen}, Geert and {Baz{\'a}n}, Juanjo and {Biswas}, Manish and {Boquien}, M{\'e}d{\'e}ric and {Burke}, D.~J. and {Cara}, Daria and {Cara}, Mihai and {Conroy}, Kyle E. and {Conseil}, Simon and {Craig}, Matthew W. and {Cross}, Robert M. and {Cruz}, Kelle L. and {D'Eugenio}, Francesco and {Dencheva}, Nadia and {Devillepoix}, Hadrien A.~R. and {Dietrich}, J{\"o}rg P. and {Eigenbrot}, Arthur Davis and {Erben}, Thomas and {Ferreira}, Leonardo and {Foreman-Mackey}, Daniel and {Fox}, Ryan and {Freij}, Nabil and {Garg}, Suyog and {Geda}, Robel and {Glattly}, Lauren and {Gondhalekar}, Yash and {Gordon}, Karl D. and {Grant}, David and {Greenfield}, Perry and {Groener}, Austen M. and {Guest}, Steve and {Gurovich}, Sebastian and {Handberg}, Rasmus and {Hart}, Akeem and {Hatfield-Dodds}, Zac and {Homeier}, Derek and {Hosseinzadeh}, Griffin and {Jenness}, Tim and {Jones}, Craig K. and {Joseph}, Prajwel and {Kalmbach}, J. Bryce and {Karamehmetoglu}, Emir and {Ka{\l}uszy{\'n}ski}, Miko{\l}aj and {Kelley}, Michael S.~P. and {Kern}, Nicholas and {Kerzendorf}, Wolfgang E. and {Koch}, Eric W. and {Kulumani}, Shankar and {Lee}, Antony and {Ly}, Chun and {Ma}, Zhiyuan and {MacBride}, Conor and {Maljaars}, Jakob M. and {Muna}, Demitri and {Murphy}, N.~A. and {Norman}, Henrik and {O'Steen}, Richard and {Oman}, Kyle A. and {Pacifici}, Camilla and {Pascual}, Sergio and {Pascual-Granado}, J. and {Patil}, Rohit R. and {Perren}, Gabriel I. and {Pickering}, Timothy E. and {Rastogi}, Tanuj and {Roulston}, Benjamin R. and {Ryan}, Daniel F. and {Rykoff}, Eli S. and {Sabater}, Jose and {Sakurikar}, Parikshit and {Salgado}, Jes{\'u}s and {Sanghi}, Aniket and {Saunders}, Nicholas and {Savchenko}, Volodymyr and {Schwardt}, Ludwig and {Seifert-Eckert}, Michael and {Shih}, Albert Y. and {Jain}, Anany Shrey and {Shukla}, Gyanendra and {Sick}, Jonathan and {Simpson}, Chris and {Singanamalla}, Sudheesh and {Singer}, Leo P. and {Singhal}, Jaladh and {Sinha}, Manodeep and {Sip{\H{o}}cz}, Brigitta M. and {Spitler}, Lee R. and {Stansby}, David and {Streicher}, Ole and {{\v{S}}umak}, Jani and {Swinbank}, John D. and {Taranu}, Dan S. and {Tewary}, Nikita and {Tremblay}, Grant R. and {de Val-Borro}, Miguel and {Van Kooten}, Samuel J. and {Vasovi{\'c}}, Zlatan and {Verma}, Shresth and {de Miranda Cardoso}, Jos{\'e} Vin{\'\i}cius and {Williams}, Peter K.~G. and {Wilson}, Tom J. and {Winkel}, Benjamin and {Wood-Vasey}, W.~M. and {Xue}, Rui and {Yoachim}, Peter and {Zhang}, Chen and {Zonca}, Andrea and {Astropy Project Contributors}},
        title = "{The Astropy Project: Sustaining and Growing a Community-oriented Open-source Project and the Latest Major Release (v5.0) of the Core Package}",
      journal = {\apj},
         year = 2022,
        month = aug,
       volume = {935},
       number = {2},
          eid = {167},
        pages = {167},
          doi = {10.3847/1538-4357/ac7c74},
archivePrefix = {arXiv},
       eprint = {2206.14220},
 primaryClass = {astro-ph.IM},
       adsurl = {https://ui.adsabs.harvard.edu/abs/2022ApJ...935..167A}
}

@INPROCEEDINGS{1979BAAS...11Q.724W,
       author = {{Weaver}, T.~A. and {Woosley}, S.~E.},
        title = "{Evolution and Final Fate of 10M$_{{\ensuremath{\odot}}}$ Stars}",
    booktitle = {Bulletin of the American Astronomical Society},
         year = 1979,
       volume = {11},
        month = dec,
        pages = {724},
       adsurl = {https://ui.adsabs.harvard.edu/abs/1979BAAS...11Q.724W}
}

@ARTICLE{2015ApJ...810...34W,
       author = {{Woosley}, S.~E. and {Heger}, Alexander},
        title = "{The Remarkable Deaths of 9-11 Solar Mass Stars}",
      journal = {\apj},
         year = 2015,
        month = sep,
       volume = {810},
       number = {1},
          eid = {34},
        pages = {34},
          doi = {10.1088/0004-637X/810/1/34},
archivePrefix = {arXiv},
       eprint = {1505.06712},
 primaryClass = {astro-ph.SR},
       adsurl = {https://ui.adsabs.harvard.edu/abs/2015ApJ...810...34W}
}

@ARTICLE{2010MNRAS.405.2113D,
       author = {{Dessart}, Luc and {Livne}, Eli and {Waldman}, Roni},
        title = "{Shock-heating of stellar envelopes: a possible common mechanism at the origin of explosions and eruptions in massive stars}",
      journal = {\mnras},
         year = 2010,
        month = jul,
       volume = {405},
       number = {4},
        pages = {2113-2131},
          doi = {10.1111/j.1365-2966.2010.16626.x},
archivePrefix = {arXiv},
       eprint = {0910.3655},
 primaryClass = {astro-ph.SR},
       adsurl = {https://ui.adsabs.harvard.edu/abs/2010MNRAS.405.2113D}
}

@ARTICLE{limongi2024,
       author = {{Limongi}, Marco and {Roberti}, Lorenzo and {Chieffi}, Alessandro and {Nomoto}, Ken'ichi},
        title = "{Evolution and Final Fate of Solar Metallicity Stars in the Mass Range 7{\textendash}15 M $_{{\ensuremath{\odot}}}$. I. The Transition from Asymptotic Giant Branch to Super-AGB Stars, Electron Capture, and Core-collapse Supernova Progenitors}",
      journal = {\apjs},
         year = 2024,
        month = feb,
       volume = {270},
       number = {2},
          eid = {29},
        pages = {29},
          doi = {10.3847/1538-4365/ad12c1},
archivePrefix = {arXiv},
       eprint = {2312.00107},
 primaryClass = {astro-ph.SR},
       adsurl = {https://ui.adsabs.harvard.edu/abs/2024ApJS..270...29L}
}

@ARTICLE{yaron2017,
       author = {{Yaron}, O. and {Perley}, D.~A. and {Gal-Yam}, A. and {Groh}, J.~H. and {Horesh}, A. and {Ofek}, E.~O. and {Kulkarni}, S.~R. and {Sollerman}, J. and {Fransson}, C. and {Rubin}, A. and {Szabo}, P. and {Sapir}, N. and {Taddia}, F. and {Cenko}, S.~B. and {Valenti}, S. and {Arcavi}, I. and {Howell}, D.~A. and {Kasliwal}, M.~M. and {Vreeswijk}, P.~M. and {Khazov}, D. and {Fox}, O.~D. and {Cao}, Y. and {Gnat}, O. and {Kelly}, P.~L. and {Nugent}, P.~E. and {Filippenko}, A.~V. and {Laher}, R.~R. and {Wozniak}, P.~R. and {Lee}, W.~H. and {Rebbapragada}, U.~D. and {Maguire}, K. and {Sullivan}, M. and {Soumagnac}, M.~T.},
        title = "{Confined dense circumstellar material surrounding a regular type II supernova}",
      journal = {Nature Physics},
         year = 2017,
        month = feb,
       volume = {13},
       number = {5},
        pages = {510-517},
          doi = {10.1038/nphys4025},
archivePrefix = {arXiv},
       eprint = {1701.02596},
 primaryClass = {astro-ph.HE},
       adsurl = {https://ui.adsabs.harvard.edu/abs/2017NatPh..13..510Y}
}

@ARTICLE{langer2012,
       author = {{Langer}, N.},
        title = "{Presupernova Evolution of Massive Single and Binary Stars}",
      journal = {\araa},
         year = 2012,
        month = sep,
       volume = {50},
        pages = {107-164},
          doi = {10.1146/annurev-astro-081811-125534},
archivePrefix = {arXiv},
       eprint = {1206.5443},
 primaryClass = {astro-ph.SR},
       adsurl = {https://ui.adsabs.harvard.edu/abs/2012ARA&A..50..107L}
}

@BOOK{kippenhahn2012,
       author = {{Kippenhahn}, Rudolf and {Weigert}, Alfred and {Weiss}, Achim},
        title = "{Stellar Structure and Evolution}",
         year = 2013,
          doi = {10.1007/978-3-642-30304-3},
       adsurl = {https://ui.adsabs.harvard.edu/abs/2013sse..book.....K}
}

@ARTICLE{2020ApJ...891L..32M,
       author = {{Morozova}, Viktoriya and {Piro}, Anthony L. and {Fuller}, Jim and {Van Dyk}, Schuyler D.},
        title = "{The Influence of Late-stage Nuclear Burning on Red Supergiant Supernova Light Curves}",
      journal = {\apjl},
         year = 2020,
        month = mar,
       volume = {891},
       number = {2},
          eid = {L32},
        pages = {L32},
          doi = {10.3847/2041-8213/ab77c8},
archivePrefix = {arXiv},
       eprint = {1912.10050},
 primaryClass = {astro-ph.HE},
       adsurl = {https://ui.adsabs.harvard.edu/abs/2020ApJ...891L..32M}
}

@ARTICLE{2017MNRAS.469L.108M,
       author = {{Moriya}, Takashi J. and {Yoon}, Sung-Chul and {Gr{\"a}fener}, G{\"o}tz and {Blinnikov}, Sergei I.},
        title = "{Immediate dense circumstellar environment of supernova progenitors caused by wind acceleration: its effect on supernova light curves}",
      journal = {\mnras},
         year = 2017,
        month = jul,
       volume = {469},
       number = {1},
        pages = {L108-L112},
          doi = {10.1093/mnrasl/slx056},
archivePrefix = {arXiv},
       eprint = {1703.03084},
 primaryClass = {astro-ph.HE},
       adsurl = {https://ui.adsabs.harvard.edu/abs/2017MNRAS.469L.108M}
}

@ARTICLE{2021ApJ...906....1S,
       author = {{Soker}, Noam},
        title = "{A Pre-explosion Extended Effervescent Zone around Core-collapse Supernova Progenitors}",
      journal = {\apj},
         year = 2021,
        month = jan,
       volume = {906},
       number = {1},
          eid = {1},
        pages = {1},
          doi = {10.3847/1538-4357/abca8f},
archivePrefix = {arXiv},
       eprint = {2010.01531},
 primaryClass = {astro-ph.HE},
       adsurl = {https://ui.adsabs.harvard.edu/abs/2021ApJ...906....1S}
}

@ARTICLE{2024OJAp....7E..47F,
       author = {{Fuller}, Jim and {Tsuna}, Daichi},
        title = "{Boil-off of red supergiants: mass loss and type II-P supernovae}",
      journal = {The Open Journal of Astrophysics},
         year = 2024,
        month = jun,
       volume = {7},
          eid = {47},
        pages = {47},
          doi = {10.33232/001c.120130},
archivePrefix = {arXiv},
       eprint = {2405.21049},
 primaryClass = {astro-ph.SR},
       adsurl = {https://ui.adsabs.harvard.edu/abs/2024OJAp....7E..47F}
}

@INPROCEEDINGS{2010ASPC..425..181B,
       author = {{Bennett}, P.~D.},
        title = "{Chromospheres and Winds of Red Supergiants: An Empirical Look at Outer Atmospheric Structure}",
    booktitle = {Hot and Cool: Bridging Gaps in Massive Star Evolution},
         year = 2010,
       editor = {{Leitherer}, C. and {Bennett}, P.~D. and {Morris}, P.~W. and {Van Loon}, J. Th.},
       series = {Astronomical Society of the Pacific Conference Series},
       volume = {425},
        month = jun,
        pages = {181},
          doi = {10.48550/arXiv.1004.1853},
archivePrefix = {arXiv},
       eprint = {1004.1853},
 primaryClass = {astro-ph.SR},
       adsurl = {https://ui.adsabs.harvard.edu/abs/2010ASPC..425..181B}
}

@ARTICLE{2018NatAs...2..808F,
       author = {{F{\"o}rster}, F. and {Moriya}, T.~J. and {Maureira}, J.~C. and {Anderson}, J.~P. and {Blinnikov}, S. and {Bufano}, F. and {Cabrera-Vives}, G. and {Clocchiatti}, A. and {de Jaeger}, T. and {Est{\'e}vez}, P.~A. and {Galbany}, L. and {Gonz{\'a}lez-Gait{\'a}n}, S. and {Gr{\"a}fener}, G. and {Hamuy}, M. and {Hsiao}, E.~Y. and {Huentelemu}, P. and {Huijse}, P. and {Kuncarayakti}, H. and {Mart{\'\i}nez}, J. and {Medina}, G. and {Olivares E.}, F. and {Pignata}, G. and {Razza}, A. and {Reyes}, I. and {San Mart{\'\i}n}, J. and {Smith}, R.~C. and {Vera}, E. and {Vivas}, A.~K. and {de Ugarte Postigo}, A. and {Yoon}, S. -C. and {Ashall}, C. and {Fraser}, M. and {Gal-Yam}, A. and {Kankare}, E. and {Le Guillou}, L. and {Mazzali}, P.~A. and {Walton}, N.~A. and {Young}, D.~R.},
        title = "{The delay of shock breakout due to circumstellar material evident in most type II supernovae}",
      journal = {Nature Astronomy},
         year = 2018,
        month = sep,
       volume = {2},
        pages = {808},
          doi = {10.1038/s41550-018-0563-4},
archivePrefix = {arXiv},
       eprint = {1809.06379},
 primaryClass = {astro-ph.HE},
       adsurl = {https://ui.adsabs.harvard.edu/abs/2018NatAs...2..808F}
}

@ARTICLE{2012MNRAS.423L..92Q,
       author = {{Quataert}, E. and {Shiode}, J.},
        title = "{Wave-driven mass loss in the last year of stellar evolution: setting the stage for the most luminous core-collapse supernovae}",
      journal = {\mnras},
         year = 2012,
        month = jun,
       volume = {423},
       number = {1},
        pages = {L92-L96},
          doi = {10.1111/j.1745-3933.2012.01264.x},
archivePrefix = {arXiv},
       eprint = {1202.5036},
 primaryClass = {astro-ph.SR},
       adsurl = {https://ui.adsabs.harvard.edu/abs/2012MNRAS.423L..92Q}
}

@ARTICLE{2014ApJ...785...82S,
       author = {{Smith}, Nathan and {Arnett}, W. David},
        title = "{Preparing for an Explosion: Hydrodynamic Instabilities and Turbulence in Presupernovae}",
      journal = {\apj},
         year = 2014,
        month = apr,
       volume = {785},
       number = {2},
          eid = {82},
        pages = {82},
          doi = {10.1088/0004-637X/785/2/82},
archivePrefix = {arXiv},
       eprint = {1307.5035},
 primaryClass = {astro-ph.SR},
       adsurl = {https://ui.adsabs.harvard.edu/abs/2014ApJ...785...82S}
}

@ARTICLE{2020A&A...635A.127K,
       author = {{Kuriyama}, Naoto and {Shigeyama}, Toshikazu},
        title = "{Radiation hydrodynamical simulations of eruptive mass loss from progenitors of Type Ibn/IIn supernovae}",
      journal = {\aap},
         year = 2020,
        month = mar,
       volume = {635},
          eid = {A127},
        pages = {A127},
          doi = {10.1051/0004-6361/201937226},
archivePrefix = {arXiv},
       eprint = {1912.09738},
 primaryClass = {astro-ph.SR},
       adsurl = {https://ui.adsabs.harvard.edu/abs/2020A&A...635A.127K}
}

@ARTICLE{2020ApJ...900...99L,
       author = {{Leung}, Shing-Chi and {Fuller}, Jim},
        title = "{Hydrodynamic Simulations of Pre-supernova Outbursts in Red Supergiants: Asphericity and Mass Loss}",
      journal = {\apj},
         year = 2020,
        month = sep,
       volume = {900},
       number = {2},
          eid = {99},
        pages = {99},
          doi = {10.3847/1538-4357/abac5d},
archivePrefix = {arXiv},
       eprint = {2007.11712},
 primaryClass = {astro-ph.HE},
       adsurl = {https://ui.adsabs.harvard.edu/abs/2020ApJ...900...99L}
}

@ARTICLE{2021ApJ...908...23M,
       author = {{Matzner}, Christopher D. and {Ro}, Stephen},
        title = "{Wave-driven Shocks in Stellar Outbursts: Dynamics, Envelope Heating, and Nascent Blast Waves}",
      journal = {\apj},
         year = 2021,
        month = feb,
       volume = {908},
       number = {1},
          eid = {23},
        pages = {23},
          doi = {10.3847/1538-4357/abd03b},
archivePrefix = {arXiv},
       eprint = {2011.08861},
 primaryClass = {astro-ph.SR},
       adsurl = {https://ui.adsabs.harvard.edu/abs/2021ApJ...908...23M}
}

@ARTICLE{2024ApJ...974..270C,
       author = {{Cheng}, Shelley J. and {Goldberg}, Jared A. and {Cantiello}, Matteo and {Bauer}, Evan B. and {Renzo}, Mathieu and {Conroy}, Charlie},
        title = "{A Model for Eruptive Mass Loss in Massive Stars}",
      journal = {\apj},
         year = 2024,
        month = oct,
       volume = {974},
       number = {2},
          eid = {270},
        pages = {270},
          doi = {10.3847/1538-4357/ad701e},
archivePrefix = {arXiv},
       eprint = {2405.12274},
 primaryClass = {astro-ph.SR},
       adsurl = {https://ui.adsabs.harvard.edu/abs/2024ApJ...974..270C}
}

@ARTICLE{2023ApJ...945..104T,
       author = {{Tsuna}, Daichi and {Takei}, Yuki and {Shigeyama}, Toshikazu},
        title = "{Precursors of Supernovae from Mass Eruption: Prospects for Early Warning of Nearby Core-collapse Supernovae}",
      journal = {\apj},
         year = 2023,
        month = mar,
       volume = {945},
       number = {2},
          eid = {104},
        pages = {104},
          doi = {10.3847/1538-4357/acbbc6},
archivePrefix = {arXiv},
       eprint = {2208.08256},
 primaryClass = {astro-ph.HE},
       adsurl = {https://ui.adsabs.harvard.edu/abs/2023ApJ...945..104T}
}

@ARTICLE{2022ApJ...936..114M,
       author = {{Matsumoto}, Tatsuya and {Metzger}, Brian D.},
        title = "{Supernova Precursor Emission and the Origin of Pre-explosion Stellar Mass Loss}",
      journal = {\apj},
         year = 2022,
        month = sep,
       volume = {936},
       number = {2},
          eid = {114},
        pages = {114},
          doi = {10.3847/1538-4357/ac892c},
archivePrefix = {arXiv},
       eprint = {2206.08377},
 primaryClass = {astro-ph.HE},
       adsurl = {https://ui.adsabs.harvard.edu/abs/2022ApJ...936..114M}
}

@ARTICLE{2020MNRAS.493..468D,
       author = {{Davies}, Ben and {Beasor}, Emma R.},
        title = "{The `red supergiant problem': the upper luminosity boundary of Type II supernova progenitors}",
      journal = {\mnras},
         year = 2020,
        month = mar,
       volume = {493},
       number = {1},
        pages = {468-476},
          doi = {10.1093/mnras/staa174},
archivePrefix = {arXiv},
       eprint = {2001.06020},
 primaryClass = {astro-ph.SR},
       adsurl = {https://ui.adsabs.harvard.edu/abs/2020MNRAS.493..468D}
}

@ARTICLE{2009CBET.1928....1M,
       author = {{Maza}, J. and {Hamuy}, M. and {Antezana}, R. and {Gonzalez}, L. and {Lopez}, P. and {Silva}, S. and {Folatelli}, G. and {Iturra}, D. and {Cartier}, R. and {Forster}, F. and {Marchi}, S. and {Rojas}, A. and {Pignata}, G. and {Conuel}, B. and {Reichart}, D. and {Ivarsen}, K. and {Haislip}, J. and {Crain}, A. and {Foster}, D. and {Nysewander}, M. and {Lacluyze}, A.},
        title = "{Supernova 2009ip in NGC 7259}",
      journal = {Central Bureau Electronic Telegrams},
         year = 2009,
        month = aug,
       volume = {1928},
        pages = {1},
       adsurl = {https://ui.adsabs.harvard.edu/abs/2009CBET.1928....1M}
}

@ARTICLE{2013MNRAS.430.1801M,
       author = {{Mauerhan}, Jon C. and {Smith}, Nathan and {Filippenko}, Alexei V. and {Blanchard}, Kyle B. and {Blanchard}, Peter K. and {Casper}, Chadwick F.~E. and {Cenko}, S. Bradley and {Clubb}, Kelsey I. and {Cohen}, Daniel P. and {Fuller}, Kiera L. and {Li}, Gary Z. and {Silverman}, Jeffrey M.},
        title = "{The unprecedented 2012 outburst of SN 2009ip: a luminous blue variable star becomes a true supernova}",
      journal = {\mnras},
         year = 2013,
        month = apr,
       volume = {430},
       number = {3},
        pages = {1801-1810},
          doi = {10.1093/mnras/stt009},
archivePrefix = {arXiv},
       eprint = {1209.6320},
 primaryClass = {astro-ph.SR},
       adsurl = {https://ui.adsabs.harvard.edu/abs/2013MNRAS.430.1801M}
}

@ARTICLE{2014ApJ...789..104O,
       author = {{Ofek}, Eran O. and {Sullivan}, Mark and {Shaviv}, Nir J. and {Steinbok}, Aviram and {Arcavi}, Iair and {Gal-Yam}, Avishay and {Tal}, David and {Kulkarni}, Shrinivas R. and {Nugent}, Peter E. and {Ben-Ami}, Sagi and {Kasliwal}, Mansi M. and {Cenko}, S. Bradley and {Laher}, Russ and {Surace}, Jason and {Bloom}, Joshua S. and {Filippenko}, Alexei V. and {Silverman}, Jeffrey M. and {Yaron}, Ofer},
        title = "{Precursors Prior to Type IIn Supernova Explosions are Common: Precursor Rates, Properties, and Correlations}",
      journal = {\apj},
         year = 2014,
        month = jul,
       volume = {789},
       number = {2},
          eid = {104},
        pages = {104},
          doi = {10.1088/0004-637X/789/2/104},
archivePrefix = {arXiv},
       eprint = {1401.5468},
 primaryClass = {astro-ph.HE},
       adsurl = {https://ui.adsabs.harvard.edu/abs/2014ApJ...789..104O}
}

@ARTICLE{2021ApJ...907...99S,
       author = {{Strotjohann}, Nora L. and {Ofek}, Eran O. and {Gal-Yam}, Avishay and {Bruch}, Rachel and {Schulze}, Steve and {Shaviv}, Nir and {Sollerman}, Jesper and {Filippenko}, Alexei V. and {Yaron}, Ofer and {Fremling}, Christoffer and {Nordin}, Jakob and {Kool}, Erik C. and {Perley}, Dan A. and {Ho}, Anna Y.~Q. and {Yang}, Yi and {Yao}, Yuhan and {Soumagnac}, Maayane T. and {Graham}, Melissa L. and {Barbarino}, Cristina and {Tartaglia}, Leonardo and {De}, Kishalay and {Goldstein}, Daniel A. and {Cook}, David O. and {Brink}, Thomas G. and {Taggart}, Kirsty and {Yan}, Lin and {Lunnan}, Ragnhild and {Kasliwal}, Mansi and {Kulkarni}, Shri R. and {Nugent}, Peter E. and {Masci}, Frank J. and {Rosnet}, Philippe and {Adams}, Scott M. and {Andreoni}, Igor and {Bagdasaryan}, Ashot and {Bellm}, Eric C. and {Burdge}, Kevin and {Duev}, Dmitry A. and {Dugas}, Alison and {Frederick}, Sara and {Goldwasser}, Samantha and {Hankins}, Matthew and {Irani}, Ido and {Karambelkar}, Viraj and {Kupfer}, Thomas and {Liang}, Jingyi and {Neill}, James D. and {Porter}, Michael and {Riddle}, Reed L. and {Sharma}, Yashvi and {Short}, Phil and {Taddia}, Francesco and {Tzanidakis}, Anastasios and {van Roestel}, Jan and {Walters}, Richard and {Zhuang}, Zhuyun},
        title = "{Bright, Months-long Stellar Outbursts Announce the Explosion of Interaction-powered Supernovae}",
      journal = {\apj},
         year = 2021,
        month = feb,
       volume = {907},
       number = {2},
          eid = {99},
        pages = {99},
          doi = {10.3847/1538-4357/abd032},
archivePrefix = {arXiv},
       eprint = {2010.11196},
 primaryClass = {astro-ph.HE},
       adsurl = {https://ui.adsabs.harvard.edu/abs/2021ApJ...907...99S}
}

@ARTICLE{1955ApJ...121..161S,
       author = {{Salpeter}, Edwin E.},
        title = "{The Luminosity Function and Stellar Evolution.}",
      journal = {\apj},
         year = 1955,
        month = jan,
       volume = {121},
        pages = {161},
          doi = {10.1086/145971},
       adsurl = {https://ui.adsabs.harvard.edu/abs/1955ApJ...121..161S}
}

@ARTICLE{2015PASA...32...16S,
       author = {{Smartt}, S.~J.},
        title = "{Observational Constraints on the Progenitors of Core-Collapse Supernovae: The Case for Missing High-Mass Stars}",
      journal = {\pasa},
         year = 2015,
        month = apr,
       volume = {32},
          eid = {e016},
        pages = {e016},
          doi = {10.1017/pasa.2015.17},
archivePrefix = {arXiv},
       eprint = {1504.02635},
 primaryClass = {astro-ph.SR},
       adsurl = {https://ui.adsabs.harvard.edu/abs/2015PASA...32...16S}
}

@ARTICLE{2025PASP..137d4203D,
       author = {{Das}, Kaustav K. and {Kasliwal}, Mansi M. and {Fremling}, Christoffer and {Sollerman}, Jesper and {Perley}, Daniel A. and {De}, Kishalay and {Tzanidakis}, Anastasios and {Sit}, Tawny and {Adams}, Scott and {Anand}, Shreya and {Ahumuda}, Tomas and {Andreoni}, Igor and {Brennan}, Se{\'a}n and {Brink}, Thomas and {Bruch}, Rachel J. and {Chen}, Ping and {Chu}, Matthew R. and {Cook}, David O. and {Covarrubias}, Sofia and {Dahiwale}, Aishwarya and {Earley}, Nicholas and {Ho}, Anna Y.~Q. and {Gal-Yam}, Avishay and {Gangopadhyay}, Anjasha and {Hammerstein}, Erica and {Hinds}, K. -Ryan and {Karambelkar}, Viraj and {Kong}, Yihan and {Kulkarni}, S.~R. and {Jegou du Laz}, Theophile and {Liu}, Chang and {Meynardie}, William and {Miller}, Adam A. and {Nir}, Guy and {Patra}, Kishore C. and {Pessi}, Priscila J. and {Rich}, R. Michael and {Rehemtulla}, Nabeel and {Rose}, Sam and {Rusholme}, Ben and {Schulze}, Steve and {Sharma}, Yashvi and {Singh}, Avinash and {Smith}, Roger and {Stein}, Robert and {Mandigo-Stoba}, Milan Sharma and {Strotjohann}, Nora L. and {Qin}, Yu-Jing and {Wise}, Jacob and {Wold}, Avery and {Yan}, Lin and {Yang}, Yi and {Yao}, Yuhan and {Zimmerman}, Erez},
        title = "{Low-luminosity Type IIP Supernovae from the Zwicky Transient Facility Census of the Local Universe. I. Luminosity Function, Volumetric Rate}",
      journal = {\pasp},
         year = 2025,
        month = apr,
       volume = {137},
       number = {4},
          eid = {044203},
        pages = {044203},
          doi = {10.1088/1538-3873/adcaeb},
archivePrefix = {arXiv},
       eprint = {2502.19493},
 primaryClass = {astro-ph.HE},
       adsurl = {https://ui.adsabs.harvard.edu/abs/2025PASP..137d4203D}
}

@ARTICLE{2025arXiv250620068D,
       author = {{Das}, Kaustav K. and {Kasliwal}, Mansi M. and {Sollerman}, Jesper and {Fremling}, Christoffer and {Moriya}, Takashi J. and {Hinds}, K-Ryan and {Perley}, Daniel A. and {Bellm}, Eric C. and {Chen}, Tracy X. and {O'Connor}, Evan P. and {Coughlin}, Michael W. and {Jacobson-Galan}, W.~V. and {Gangopadhyay}, Anjasha and {Graham}, Matthew and {Kulkarni}, S.~R. and {Purdum}, Josiah and {Sarin}, Nikhil and {Schulze}, Steve and {Singh}, Avinash and {Tsuna}, Daichi and {Wold}, Avery},
        title = "{Low-Luminosity Type IIP Supernovae from the Zwicky Transient Facility Census of the Local Universe. II: Lightcurve Analysis}",
      journal = {arXiv e-prints},
         year = 2025,
        month = jun,
          eid = {arXiv:2506.20068},
        pages = {arXiv:2506.20068},
          doi = {10.48550/arXiv.2506.20068},
archivePrefix = {arXiv},
       eprint = {2506.20068},
 primaryClass = {astro-ph.HE},
       adsurl = {https://ui.adsabs.harvard.edu/abs/2025arXiv250620068D}
}
\bibliographystyle{aasjournalv7}



\end{document}